\documentclass[final,authoryear,5p,times,twocolumn]{elsarticle}

\usepackage{epsfig}

\usepackage{amssymb}

\usepackage{lineno}
\usepackage{rotating}
\usepackage{color,soul}
\usepackage{longtable}

\usepackage[font=scriptsize,labelfont=bf]{caption}

\usepackage[bookmarks = true, bookmarksnumbered = true, pdfpagemode =None, pdfstartview = FitH, pdfpagelayout = SinglePage, colorlinks = true, urlcolor = blue, citecolor = monbleu]{hyperref}

\newcommand{\beq}{\bigskip\begin{equation}}
\newcommand{\eeq}{\bigskip\end{equation}}

\journal{Icarus}

\begin{document}

\begin{frontmatter}



\title{Moist Convection and the 2010-2011 Revival of Jupiter's South Equatorial Belt}


\author[le]{Leigh N. Fletcher}
\ead{leigh.fletcher@leicester.ac.uk}
\author[jpl]{G.S. Orton}
\author[jr]{J.H. Rogers}
\author[ox]{R.S. Giles}
\author[jpl]{A.V. Payne}
\author[ox]{P.G.J. Irwin}
\author[mv]{M. Vedovato}


%

\address[le]{Department of Physics \& Astronomy, University of Leicester, University Road, Leicester, LE1 7RH, UK}
\address[jpl]{Jet Propulsion Laboratory, California Institute of Technology, 4800 Oak Grove Drive, Pasadena, CA, 91109, USA}
\address[jr]{British Astronomical Association, Burlington House, Piccadilly, London  W1J 0DU, UK}
\address[ox]{Atmospheric, Oceanic \& Planetary Physics, Department of Physics, University of Oxford, Clarendon Laboratory, Parks Road, Oxford, OX1 3PU, UK}
\address[mv]{JUPOS Team, Unione Astrofili Italiani}




\begin{abstract}
The transformation of Jupiter's South Equatorial Belt (SEB) from its faded, whitened state in 2009-2010 \citep{11fletcher_fade} to its normal brown appearance is documented via comparisons of thermal-infrared (5-20 $\mu$m) and visible-light imaging between November 2010 and November 2011.  The SEB revival consisted of convective eruptions triggered over $\sim100$ days, potentially powered by the latent heat released by the condensation of water.  The plumes rise from the water cloud base and ultimately diverge and cool in the stably-stratified upper troposphere. Thermal-IR images from the Very Large Telescope (VLT) were acquired 2 days after the SEB disturbance was first detected as a small white spot by amateur observers on November 9th 2010.  Subsequent images over several months revealed the cold, putatively anticyclonic and cloudy plume tops (area $2.5\times10^6$ km$^2$) surrounded by warm, cloud-free conditions at their peripheries due to subsidence.  The latent heating was not directly detectable in the 5-20 $\mu$m range.  The majority of the plumes erupted from a single source near $140-160^\circ$W, coincident with the remnant cyclonic circulation of a brown barge that had formed during the fade.  The warm remnant of the cyclone could still be observed in IRTF imaging 5 days before the November 9th eruption.  Additional plumes erupted from the leading edge of the central disturbance immediately east of the source, which propagated slowly eastwards to encounter the Great Red Spot.  The tropospheric plumes were sufficiently vigorous to excite stratospheric thermal waves over the SEB with a $20-30^\circ$ longitudinal wavelength and 5-6 K temperature contrasts at 5 mbar, showing a direct connection between moist convection and stratospheric wave activity. The subsidence and compressional heating of dry, unsaturated air warmed the troposphere (particularly to the northwest of the central branch of the revival) and removed the aerosols that had been responsible for the fade.  Dark, cloud-free lanes west of the plumes were the first to show the colour change, and elongated due to the zonal windshear to form the characteristic `S-shape' of the revival complex.  The aerosol-free air was redistributed and mixed throughout the SEB by the zonal flow, following a westward-moving southern branch and an eastward-moving northern branch that revived the brown colouration over $\sim200$ days.  The transition from the cool conditions of the SEBZ during the fade to the revived SEB caused a 2-4 K rise in 500-mbar temperatures (leaving a particularly warm southern SEB) and a reduction of aerosol opacity by factors of 2-3.  Newly-cleared gaps in the upper tropospheric aerosol layer appeared different in filters sensing the $\sim700$-mbar cloud deck and the 2-3 bar cloud deck, suggesting complex vertical structure in the downdrafts.  The last stage of the revival was the re-establishment of normal convective activity northwest of the GRS in September 2011, $\sim840$ days after the last occurrence in June 2009.  Moist convection may therefore play an important role in controlling the timescale and atmospheric variability during the SEB life cycle.

\end{abstract}


\end{frontmatter}


\section{Introduction}
\label{intro}

Global upheavals of Jupiter's banded structure can produce striking changes to the planet's appearance over a period of just a few months, capturing the attention of amateur and professional observers alike and prompting intense periods of scrutiny of the giant planet \citep{95rogers}.  The temporal variability of Jupiter's bright, cloudy zones and darker, cloud-free belts can reveal the atmospheric processes shaping the gas giant's typical appearance, including large-scale overturning, moist convective dynamics, and cloud condensation.  Between 2009 and 2010, Jupiter's South Equatorial Belt (SEB, between the prograde SEBn jet at $6.1^\circ$S and the retrograde SEBs jet at $17.2^\circ$S, planetocentric latitudes) underwent a dramatic change from being the darkest and broadest belt on the planet to a faded, white, zone-like state.  This signalled the start of a whitening and re-darkening cycle, known as an SEB fade and revival, that was investigated for the first time in the thermal infrared (5-25 $\mu$m) to understand the temperature and aerosol changes associated with the `disappearance' (whitening) of the belt \citep[][hereafter Paper 1]{11fletcher_fade}.  Narrow-band thermal-IR images were presented between May 2008 and September 2010, tracking the shutdown of the normal convective activity in the SEB through to the fully-faded and quiescent state in mid-2010.  

Here we present the first characterisation of thermal changes associated with a revival of the SEB.  Section \ref{obs} describes the coordinated campaign of 5-20 $\mu$m imaging necessary to track the event, utilising data from the VLT and Gemini-South in Chile; the IRTF, Subaru and Gemini-North in Hawaii; in tandem with amateur observers from around the globe.  A comprehensive observational timeline for the revival is given in Section \ref{timeline}, from the detection of SEB plumes in November 2010 to the return to `normal' SEB conditions in late 2011.  This includes temperature and aerosol measurements from inversions of the infrared data.  Section \ref{discuss} presents an analogy between the SEB revival plumes and mesoscale convective systems on Earth, using the dynamics of moist convection to explain the thermal variations observed during the SEB revival.

\subsection{Historical background}
Historical accounts of colour changes during SEB life cycles \citep{58peek, 95rogers, 96sanchez} revealed similar patterns during each fade and revival, enabling predictions for the general progression of this episodic event.  As summarised in Paper 1, the 2009-10 cycle was the fifth since the first spacecraft encounter with Jupiter (Pioneer 10 in 1973), with SEB fades and revivals occurring between 1972-1975 \citep{81orton, 95rogers}, 1989-1990 \citep{92rogers, 92satoh, 93kuehn, 94satoh}, 1992-1993 \citep{96sanchez, 97moreno} and a partial fade in 2007 \citep{07reuter, 07baines}.  The twentieth century saw fourteen such revivals, from the 1919/20 cycle - the first to be observed in detail after a quiet period of 37 years - to the 1992/93 cycle \citep{95rogers}.  The gap between the revival and the next fade can be anything between 2-14 years, making the onset of the fade difficult to predict.  

The 2009-10 fade sequence can be summarised as follows (see Table 4 of Paper 1).  The filamentary convection that normally dominates the `wake' to the northwest of the Great Red Spot had largely ceased by late May 2009, signalling the onset of a more quiescent state for the SEB.  By July a band of elevated cloud opacity filled the centre of the SEB (known as the SEB zone, or SEBZ), separating the cloud-free northern and southern components of the belt, the SEB(N) and SEB(S), respectively.  This increased cloudiness was observed at 4.8 and 8.6 $\mu$m, which sense opacity near the 2-3 bar level and 800-mbar level, respectively, and preceded the visible fade of the SEB by several months.  Using a 3-layer aerosol distribution to model visible reflectivity, \citet{12perezhoyos} confirmed that the fade was detected earliest at the deepest-sensing wavelengths, and caused a substantial increase in the optical thickness of the topmost condensation cloud ($1.0\pm0.4$ bar) and reflectivity of aerosols throughout the upper troposphere (0.24-1.4 bar).  Such a sequence hinted at a large-scale change diffusing upwards from the deeper atmosphere and affecting all longitudes simultaneously.  There appeared to be no detectable change in the cloud altitudes or cloud-tracked winds during the fade \citep{12perezhoyos}, and \citet{11fletcher_fade} speculated that enhanced vertical mixing of volatile-laden air, combined with condensation in the cool SEBZ, served to modify pre-existing aerosol layers.  The upwelling also caused the SEBZ to exhibit cool zone-like temperatures for $p>300$ mbar (i.e., within the convective region), but no SEBZ was detectable at lower pressures in the radiatively-controlled upper troposphere.  Ultimately, the change in opacity and reflectivity of the tropospheric aerosols obscured the brown chromophores of the SEB and caused the whole belt to whiten over, although we caution the reader that no spectroscopically-identifiable ammonia ice signatures \citep{02baines} have been reported during this faded state.  

By August 2009 the SEB(S) had also begun to cloud over and take on a paler appearance, and by November 2009 it was indistinguishable from the rest of the SEB.  Only the SEB(N) remained dark and cloud-free (i.e., bright at 4.8 $\mu$m) by the end of the 2009 apparition (May 2009-January 2010).  During the onset of the fade, a series of five brown barges (cyclonic, cloud-free and warm-temperature ovals, named B1 to B5) formed in the pale brown SEB(S) at progressively greater longitudes west of the Great Red Spot \citep{11rogers_21, 11fletcher_fade, 12perezhoyos}.    The first barge appeared just west of the GRS in June 2009, the fifth barge formed in August, and these cyclonic regions appeared to be an unusual new feature of the SEB faded state.  These barges were dark and conspicuous in January 2010, but these too faded and were only faintly visible as oval outlines when Jupiter reappeared from behind the Sun in April 2010.  By July 2010, when the fade was complete, the brown barges were only detectable as faint thermal signatures at 10.8 $\mu$m (sensing temperatures and ammonia gas near 500 mbar).  By this time the whole SEB was pale in colour and dark at 4.8 $\mu$m and 8.6 $\mu$m due to a two-fold increase in tropospheric mid-IR aerosol opacity since mid-2008. 

\subsection{Expectations for the revival}
Paper 1 demonstrated that the normal convective motions that typify the SEB had completely shut down by mid-2010, and suggested that this was replaced by zone-like upwelling of fresh condensables like NH$_3$ gas.  We hypothesised that removal of the white colouration would require sublimation of the ices due to a temperature rise, leading to a revival of the SEB.  But how would such a temperature rise occur? Would it be a slow or rapid change?  Historical records \citep{95rogers, 96sanchez} suggest that a single white convective plume (the SEB disturbance, or SEBD) would initiate the process by punching through the stagnant, quiescent white cloud layer.  The disturbance would then spread in a violent and eruptive manner, producing three distinct branches:  a central branch where the interior of the SEB is reviving; a northern branch moving eastward to restore the SEB(N); and a southern branch of dark spots moving westward to restore the SEB(S) \citep{95rogers}.  These three branches were all observed in visible-light imaging during the 2010-11 revival \citep{11rogers_21, 16rogers}, but how do they relate to the changing temperature field?  Could a belt-wide temperature rise precede the SEBD, gradually creating conditions suitable for the colour change \citep{81orton, 94satoh}?  Alternatively, the rising plumes could directly inject latent heat as gaseous species condense out, or large-scale sinking motions surrounding small-scale plumes could warm the atmosphere by subsidence and adiabatic warming.  Furthermore, Paper 1 suggested that the SEBD might be related to the residual cyclonic circulations (subsidence and warming) associated with the five brown barges:  the 2007 SEBD erupted from a brown spot \citep{07rogers}, and the same may have occurred in 1993 \citep{96sanchez}.  In the following sections, we present thermal-infrared imaging during the revival to shed new light on the environmental changes associated with this colour change.

\section{Observations and Spectral Inversion}
\label{obs}

Paper 1 focussed solely on thermal-IR observations of the SEB fade from the European Southern Observatory (ESO) Very Large Telescope (VLT) in Chile and the NASA Infrared Telescope Facility (IRTF) in Hawaii.  However, capturing such a rapidly evolving revival event required the use of multiple facilities in the M (4.5-5.2 $\mu$m), N (8-13 $\mu$m) and Q (17-24 $\mu$m) bands as outlined below.  Images were initially reduced using instrument-specific software provided by the observatories.  Despiking and bad pixel removal, geometrical registration by fitting the limb, cylindrical reprojection and absolute calibration were performed following techniques described in \citet{09fletcher_imaging}.  Radiometric calibration in the 7-24 $\mu$m range was achieved via scaling the central meridian radiances in a particular image to zonal mean Voyager/IRIS and Cassini/CIRS spectra convolved with the relevant filter function, avoiding the region of the SEB.  All latitudes in this study are planetocentric, all longitudes are quoted for Jupiter's System III West.  The following subsections briefly describe the seven instruments used in this work.

\begin{table*}[htdp]
\caption{Infrared observations of the SEB revival.}
\begin{center}
\begin{tabular}{|l|l|l|l|l|}
\hline
Date & Instrument & Programme ID & Comments & Figures \\
\hline
2010-11 Apparition & & & & \\
\hline
2010-11-04 & IRTF/MIRSI & 2010B010 & Central longitude before revival plume & \ref{prestorm}, \ref{montage2010} \\
2010-11-11 & VLT/VISIR & 286.C-5009 & Revival plume at 8.6 $\mu$m & \ref{montage2010} \\
2010-11-13 & IRTF/SpeX &  2010B010 & Revival plume at 5 $\mu$m & \ref{montage5um} \\
2010-11-13 & VLT/VISIR & 286.C-5009 & Non-reviving hemisphere, all filters & \ref{montage2010} \\
2010-11-16 & IRTF/NSFCAM2 & 2010B010 & Central branch at 5 $\mu$m &  \ref{montage5um} \\
2010-11-18 & Gemini-N/NIRI & GN-2010B-DD-3 & 5-$\mu$m lucky imaging &  \ref{montage5um} \\
2010-11-21 & Gemini-S/TReCS & GS-2010B-Q-8 & Acquisition imaging, central branch & \ref{montage2010} \\
2010-11-21 & Gemini-N/NIRI & GN-2010B-DD-3 & 5-$\mu$m lucky imaging &  \ref{montage5um} \\
2010-11-28 & IRTF/NSFCAM2 & 2010B010& Central branch at 5 $\mu$m &  \ref{montage5um}  \\
2010-11-30 & Gemini-N/NIRI & GN-2010B-DD-3 & 5-$\mu$m lucky imaging &  \ref{montage5um}  \\
2010-12-01 & VLT/VISIR & 286.C-5009 & Revival central branch, all filters & \ref{montage2010}, \ref{visir_label}, \ref{filters} \\
2010-12-05/06 & IRTF/MIRSI & 2010B010 & Central branch & \ref{montage2010} \\
2010-12-05 & IRTF/NSFCAM2 & 2010B010& Central branch at 5 $\mu$m &  \ref{5um_maps}, \ref{montage5um} \\
2010-12-10 & IRTF/NSFCAM2 & 2010B010& Central branch at 5 $\mu$m &  \ref{montage5um} \\
2011-01-06 & IRTF/NSFCAM2 & 2010B012& 5 $\mu$m &  \ref{montage5um} \\
2011-01-09 & IRTF/NSFCAM2 &2010B012 & 5 $\mu$m &  \ref{montage5um} \\
2011-01-16 & VLT/VISIR & 286.C-5009 & Revival central branch, all filters & \ref{visir_label},  \ref{montage2011} \\
2011-01-27 & IRTF/SpeX & 2010B012 & Mature revival at 5 $\mu$m &  \ref{montage5um}  \\
2011-01-27/31 & VLT/VISIR & 087.C-0024(A) & Revival global coverage & \ref{montage2011}, \ref{cmap2011jan} \\
2011-03-01 & IRTF/NSFCAM2 & 2011A010 & 5 $\mu$m &  \ref{montage5um} \\
\hline
2011-12 Apparition & & & & \\
\hline
2011-07-25/26 & IRTF/MIRSI & 2011A010  & Low quality, overlapped Gemini-S/TReCS & Not shown \\
2011-07-27 & Gemini-S/TReCS & GS-2011A-Q-11 & Acquisition imaging, mature revival & \ref{montage2011} \\
2011-08-27 & Subaru/COMICS & O11154 & Near-global map in all filters & \ref{cmap2011aug} \\
2011-08-29/31 & IRTF/NSFCAM2 & 2011B027 & 5 $\mu$m, Mature revival & \ref{5um_maps} \\
2011-08-30 & IRTF/MIRSI &   2011B027 & Mature revival & \ref{montage2011} \\
2011-09-13/18 & VLT/VISIR & 087.C-0024(B) & Post-revival global coverage & \ref{montage2011}, \ref{cmap2011sep} \\
2011-11-25 & Gemini-S/TReCS & GS-2011B-Q-11 & GRS convection restarted & \ref{montage2011} \\
\hline
\end{tabular}
\end{center}
\label{tab:data}
\end{table*}%

\subsection{VLT/VISIR}
The VLT Imager and Spectrometer for the mid-infrared \citep[VISIR][]{04lagage} on the 8.2-m diameter Melipal telescope has been used to study Jupiter since 2006. We observed the SEB revival on multiple occasions between November 2010 and September 2011 (Table \ref{tab:data}).  We were awarded Directors Discretionary Time (programme 286.C-5009) and regular observations (programme 087.C-0024) to extend the previous studies of the SEB fade \citep{11fletcher_fade}.  Observations were acquired in 8 discrete filters (Table \ref{tab:filters}) between 7.9 and 19.5 $\mu$m, chopping and nodding to remove the sky background.  VISIR's $256\times256$ pixel array and pixel scale of 0.127" provided a field of view of $32\times32$".  Combined with the maximum chopping amplitude of 25", it was necessary to observe northern and southern hemispheres independently (and sometimes on different nights).  A full image sequence of a single hemisphere took approximately 50 minutes to complete.  The ESO data pipeline was used for initial reduction and bad-pixel removal via its front-end interface, GASGANO.  The VISIR observations of Jupiter in September 2011 were the final ones before the instrument was removed from the telescope for refurbishment between 2012-2015.

\begin{table*}[htdp]
\caption{VLT/VISIR Filters used in this study.  Approximate peaks of the filter contribution functions are based on \citet{09fletcher_imaging}.}
\begin{center}
\begin{tabular}{|c|c|c|c|}
\hline
Name & Wavelength ($\mu$m) & Sensitivity & Approx. Pressure (mbar) \\
\hline
J7.9 & 7.90 & Stratospheric $T(p)$ & 5 \\
PAH1 & 8.59 & Aerosols and $T(p)$ & 650\\
SIV2 & 10.77 & NH$_3$, aerosols, $T(p)$ & 400 \\
NeII\_1 & 12.27 & Stratospheric $T(p)$ and C$_2$H$_6$ & 6 \\
NeII\_2 & 13.04 & Tropospheric $T(p)$ & 460 \\
Q1 & 17.65 & Tropospheric $T(p)$ & 200 \\
Q2 & 18.72 & Tropospheric $T(p)$ & 270 \\
Q3 & 19.50 & Tropospheric $T(p)$ & 400 \\
\hline
\end{tabular}
\end{center}
\label{tab:filters}
\end{table*}%

\subsection{IRTF/MIRSI}
NASA's Infrared Telescope Facility (IRTF) provided 7-24 $\mu$m imaging using the MIRSI instrument \citep[Mid Infrared Spectrometer and Imager,][]{03deutsch}, albeit at a lower diffraction-limited spatial resolution due to the smaller diameter (3 m) of the primary mirror.    MIRSI featured a $320\times240$ Si:As array with a 0.3" pixel scale over a $1.6\times1.2$ arcmin field of view, and had been observing Jupiter since 2004 \citep{10fletcher_grs}.  Unfortunately, a problem with MIRSI's electronics in late 2011, at the end of the sequence presented here, meant that the instrument was removed from the observatory pending a significant upgrade. 

\subsection{Subaru/COMICS}
The Cooled Mid-IR Camera and Spectrograph \citep[COMICS,][]{00kataza} is mounted on the 8.2-m diameter Subaru telescope operated by the Japanese National Astronomical Observatory.  Subaru/COMICS has provided observations of Jupiter since 2005, and acquired images on August 27th 2011 covering $2/3$ of Jupiter's longitudes.  COMICS features a $320\times240$ pixel array with a smaller pixel scale ($0.133$") and field of view ($42\times32$") than IRTF/MIRSI. Combined with a large chopping amplitude than VLT/VISIR, this implied that Jupiter's full disc can be observed without the need for multiple pointings.  COMICS uses a similar filter set to VISIR in Table \ref{tab:filters}.

\subsection{Gemini-S/TReCS}
The Thermal-Region Camera Spectrograph \citep[TReCS,][]{05debuizer} on Gemini-South was acquiring spectroscopic observations of Jupiter in 2010 and 2011 that will be the subject of a future publication.  However, acquisition imaging at 8.6 and 10.3 $\mu$m captured the progress of the SEB revival on November 21st 2010 and July 27th 2011.  The TReCS $320\times240$ detector and $0.09$" pixel scale provided a small field of view of $28.8\times21.6$" which, when combined with the small chopping amplitude of 15", means that the Jupiter images are often obscured by the negative chopped beam.  Although attempts were made to remove this, the resulting images are generally of a low quality.  TReCS was retired from Gemini-South in early 2012 to make way for new instrumentation.  

\subsection{Gemini-N/NIRI}
Director's discretionary time on the Near InfraRed Imager and Spectrometer \citep[NIRI,][]{03hodapp} on Gemini-North was awarded to study the onset of the SEB revival, and successfully captured the central branch on three dates in November 2010.  Each observation consisted of hundreds of short 0.2-second exposures at 4.7 $\mu$m without the adaptive optics system, selecting only the best frames (i.e., those with the least blurring due to atmospheric seeing) for coaddition.  This technique, sometimes referred to as `lucky imaging' \citep[e.g.,][]{14mousis_proam}, permits spatial resolutions approaching diffraction-limited seeing (0.15" from an 8.1-m observatory at 4.7 $\mu$m).  NIRI's $1024\times1024$ pixel InSb array, combined with the $f/32$ camera providing a 0.022" pixel scale, limits the field of view to $22\times22$".  This was mosaicked across the disc on each of the three dates, but only those positions showing the SEB revival are considered in this study.  These data were not absolutely calibrated, but the fluxes were scaled to match those of our IRTF observations (see next section).

\subsection{IRTF/NSFCAM2 and SpeX}
The 4.7-$\mu$m observations from Gemini-N/NIRI were supplemented by IRTF observations with lower spatial resolution but better time coverage throughout the fade/revival cycle.  We primarily used the NSFCAM2 instrument \citep{94shure} to extend the time sequence shown in Fig. 7 of Paper 1, monitoring the changing 4.7-$\mu$m emission.  NSFCAM2 had a $2048\times2048$ pixel array with a 0.04" pixel scale, equating to a large $81\times81$" field of view for efficient mapping of Jupiter's full disc.  NSFCAM2 was removed from the telescope in early 2012 for the installation of a new infrared detector, so the SEB revival sequence was among the last observations acquired by this facility with the old array.  Standard stars were not systematically acquired to calibrate the NSFCAM2 observations, so we adopted the relative radiometric calibration technique of \citet{98ortiz} from their study of NEB hotspots from IRTF/NSFCAM data between 1993-1997.  Finally, the NSFCAM2 images were supplemented by guide-camera observations from the SpeX instrument \citep{03rayner}.  These used a $512\times512$ pixel array with a larger 0.12" pixel scale ($60\times60$" field of view), resulting in lower-quality imaging than NSFCAM2.  Nevertheless, they are useful for extending the time series of the 4.8-$\mu$m observations.

\subsection{Amateur Imaging}
The thermal imaging from professional telescopes is supplemented throughout this article by visible-light images from numerous amateur observers.  Images were submitted to the JUPOS project\footnote{\mbox{http://jupos.org}}, British Astronomical Association\footnote{\mbox{www.britastro.org/section\_front/15}} and ALPO-Japan\footnote{\mbox{http://alpo-j.asahikawa-med.ac.jp}}.  Some are also available via PVOL (Planetary Virtual Observatory and Laboratory\footnote{\mbox{www.pvol.ehu.es}}).  These three-colour images are typically obtained from 20-41 cm aperture telescopes using the webcam-image-stacking technique \citep{14mousis_proam, 15kardasis}.  Images were geometrically registered and converted to cylindrical maps using the WinJUPOS program (created by G. Hahn), which is available at \mbox{http://jupos.org}.  Limb darkening is empirically corrected for using best-fit curves. The names of the observers and observation times are described in the image captions.

\subsection{Spectral Inversion}
Cylindrical radiance maps in VISIR N- and Q-band filters were used to form crude spectral image cubes for analysis using an optimal estimation retrieval algorithm, NEMESIS \citep{08irwin}, following identical procedures outlined in Paper 1.  The residual between the data and the spectral forward model was minimised whilst ensuring smooth and physically-plausible vertical profiles using previously derived temperature and chemical distributions as priors.  Although better suited to spectroscopy, this spectral inversion algorithm was applied to photometric imaging by \citet{09fletcher_imaging} and can be used to map Jupiter's tropospheric and stratospheric temperatures, along with estimates of the spatial distribution of mid-infrared aerosol opacity.  The reader is referred to \citet{09fletcher_imaging} for a description of the sources of spectral line data, the conversion of line data to $k$-distributions for rapid spectral inversion, and the inputs to Jupiter's \textit{a priori} atmospheric structure.  

Temperatures in the troposphere and stratosphere were defined on a grid of 80 layers between 10 bar and 0.1 mbar and the full $T(p)$ is retrieved at every location.  Retrievals of aerosol opacity, NH$_3$ and C$_2$H$_6$ simply use a scaling of (i) a compact cloud with a base at 800 mbar and extinction cross-section determined by 10-$\mu$m radius NH$_3$ ice \citep{09fletcher_ph3}; (ii) an NH$_3$ profile well-mixed at 100 ppm for $p>800$ mbar and declining at higher altitudes due to both condensation and photolysis \citep[e.g.,][]{06achterberg}; and (iii) a C$_2$H$_6$ profile increasing with height in the stratosphere following \citet{07nixon}.  With only eight photometric points defining each spectrum, this inversion suffers from low vertical resolution and degeneracies between temperature, aerosols and composition (particularly for $p>400$ mbar) that manifest as large uncertainties on retrieved quantities \citep{09fletcher_imaging}.  As a result, no variability of ammonia or ethane was detected outside of the uncertainties on the spectral inversion.  The spatial and temporal variability of temperatures and aerosol opacity during the SEB revival sequence will be discussed below.  

\section{Results:  The SEB Revival Timeline}
\label{timeline}

Interpreting such a complex, evolving phenomenon from a series of incomplete infrared snapshots was a considerable challenge.  Comprehensive reports on the SEB revival have been assembled from amateur imaging \citep[e.g.,][]{11rogers_21, 11rogers_24, 15rogers, 16rogers}, and we summarise those points directly related to the infrared imaging in the following sections. This section is organised as follows:  Section \ref{seq0} describes the initial disturbance; Section \ref{seq1} describes the early evolution of the central branch in November 2010; Section \ref{seq2} presents our first multi-spectral views of the revival complex in December 2010; the development of the northern and southern branches is revealed in Section \ref{seq3}, including the production of a stratospheric wave above the reviving SEB in Section \ref{stratwave}; and Section \ref{seq4} describes the late stages of the revival and the return to `normal' conditions in July-September 2011.  A summary of the thermal and aerosol changes is described in Section \ref{summary}, and the key events during the twelve months of the SEB revival are summarised in Table \ref{tab:timeline}.

\begin{figure*}
\begin{centering}
\centerline{\includegraphics[angle=0,scale=.90]{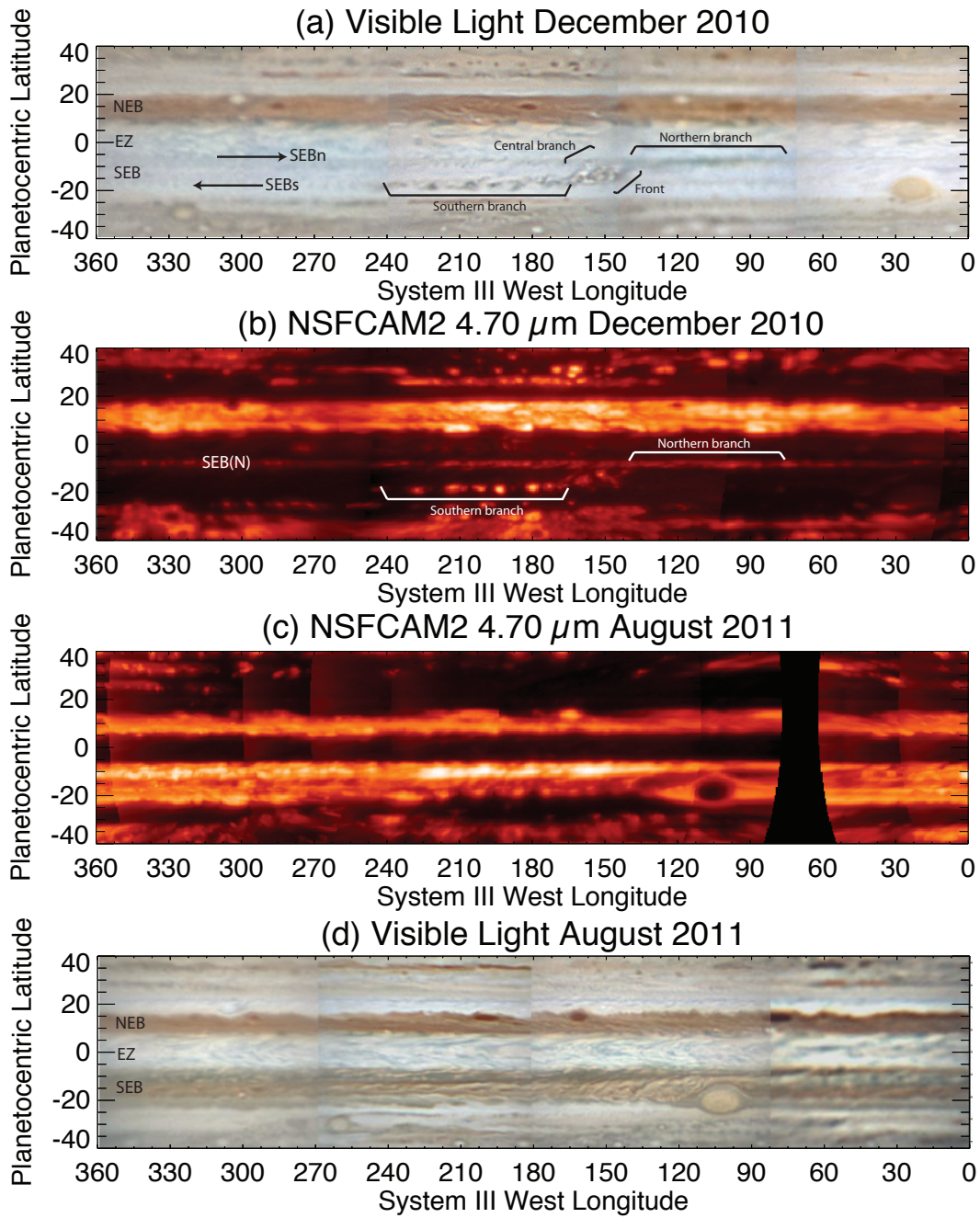}}
\caption{Global maps of Jupiter in visible light and at 5 $\mu$m from NSFCAM2, assembled over multiple nights in 2010 and 2011, showing the SEB in the middle of the revival and after the revival was complete.  The prominent features of the revival (the northern, central and southern branches), as well as the directions of the zonal jets (the SEBn and SEBs) that border the SEB, are labelled.  The visible light map in panel (a) was assembled by M. Vedovato from images between December 4th-5th by T. Kumamori, C. Go, D. Parker and K. Yunoki. The 5-$\mu$m image in panel (b) was created from images on November 27th, December 5th and 6th 2010.  The 5-$\mu$m image in panel (c) was created from images acquired on August 29th, 31st and September 1st 2011.  The visible light map in panel (d) was assembled by M. Vedovato from images between August 29th-30th by M. Jacquesson, C. Go and A. Wesley. }
\label{5um_maps}
\end{centering}
\end{figure*}

\begin{figure*}
\begin{centering}
\centerline{\includegraphics[angle=0,scale=.80]{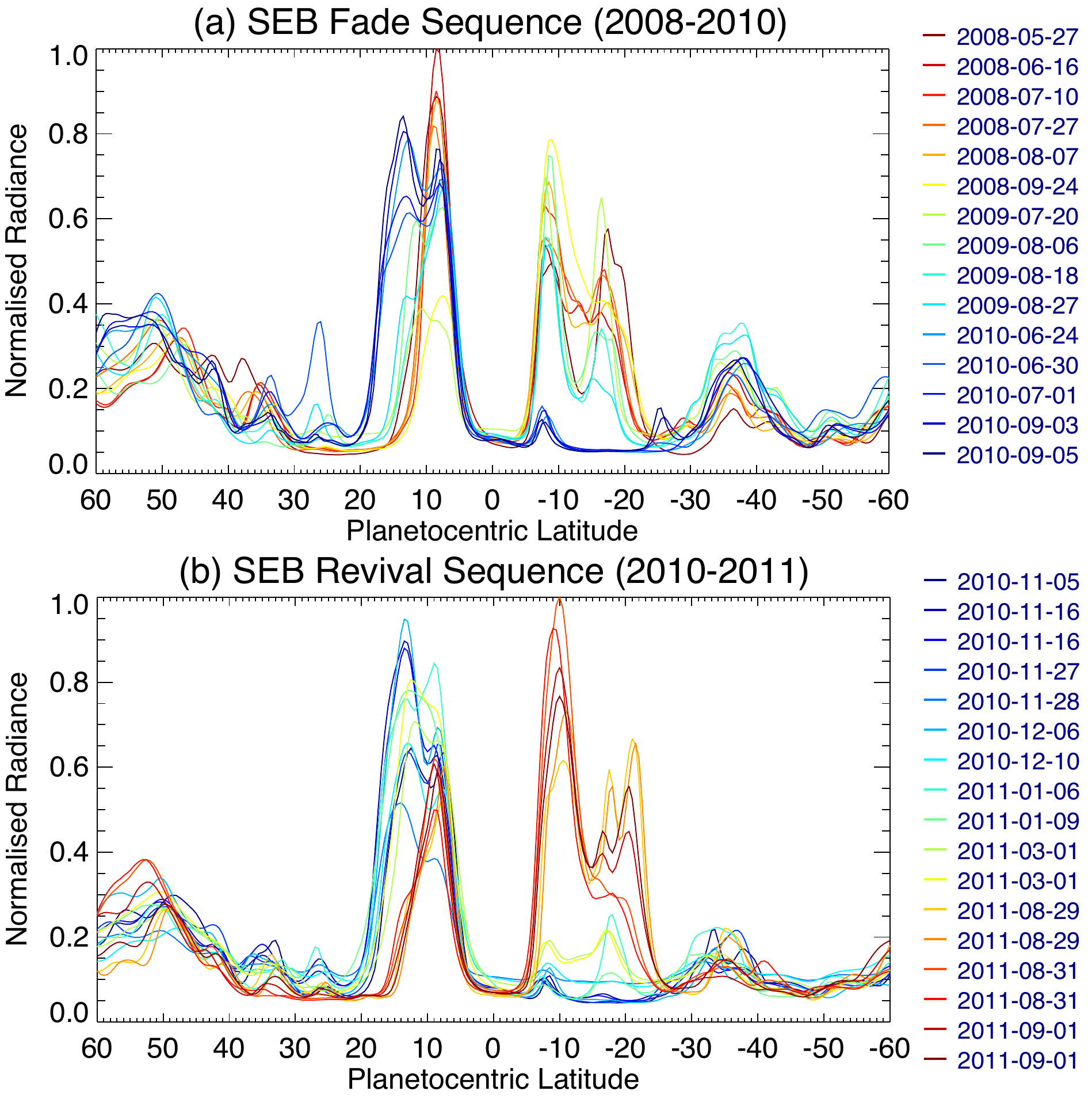}}
\caption{Central-meridian brightnesses extracted from all NSFCAM2 images at 4.7 $\mu$m during the SEB fade and revival sequence.  Radiances were averaged within $\pm50^\circ$ longitude of the central meridian for each image.  The top panel is identical to that shown by \citet{11fletcher_fade}, the bottom panel shows the new data between November 2010 and September 2011. In the absence of an absolute radiometric calibration, all radiance values have been normalised so that the equatorial brightness remains approximately unchanged.  In addition to the recovery of brightness over the SEB, an expansion and contraction event can be observed in the NEB between 2009-2011. }
\label{5um_cmerid}
\end{centering}
\end{figure*}

Before exploring the details of the revival, NSFCAM2 5-$\mu$m observations in Figs. \ref{5um_maps}  and \ref{5um_cmerid} reveal the general structure and consequences of this event.  Fig. \ref{5um_maps} compares visible and 5-$\mu$m maps mid-revival (December 2010) and post-revival (August 2011), showing the re-establishment of the typical appearance of the SEB, both in visible light (dark brown coloration) and in the infrared (bright at 5 $\mu$m, indicating a clearing of cloud opacity).  The complete clouding-over of the SEB in late 2010 made the belt almost indistinguishable from the equatorial zone (EZ) and the South Tropical Zone (STropZ), with the exception of the narrow lane of emission from the northern edge of the SEB near $7^\circ$S.  Furthermore, the Great Red Spot appeared visibly red and isolated in the faded belt and indistinguishable at 5 $\mu$m (Fig. \ref{5um_maps}a-b), whereas its periphery was once again cloud free and bright by August 2011 (Fig. \ref{5um_maps}c).  Fig. \ref{5um_cmerid} shows the evolution of 5-$\mu$m emission throughout the fade and revival sequence (2008-2011)\footnote{Figs. \ref{5um_maps} and \ref{5um_cmerid} also show an `expansion event' of the NEB, which increased in latitudinal extent in 2009 before returning to a narrower state in 2011.  These NEB expansion events occur once every 3-5 years and will be the subject of a future study.}.  Although the revival was a complex function of longitude, the net effect was a ten-fold increase in the emission at 5-$\mu$m over a period of a few months.

\begin{table*}[htdp]
\caption{Timeline of events in the 2010-2011 Revival.  For a companion table covering the fade, please see \citet{11fletcher_fade}.}
\begin{center}
\begin{tabular}{|l|p{10cm}|p{5cm}|}
\hline
\textbf{Time} & \textbf{Event} & \textbf{Possible Implication}  \\
\hline

2010-Jul-Nov & SEB fully faded \citep{11fletcher_fade}.  SEB(S) (narrow and undulating) and SEB(N) are detectable at 8.6 $\mu$m, barges cannot be distinctly observed (except at 10.8 $\mu$m); central SEB covered by high opacity cloud. & SEB fade complete.\\

2010-Sep-21 & Jupiter at opposition & Centre of 2010/11 apparition. \\

2010-Nov-04 & Brown barge B2 still faintly visible in IRTF/MIRSI imaging & \\

2010-Nov-09 & C. Go reports first evidence of SEB disturbance (WS1) and revival & Revival of SEB begins. \\

2010-Nov-09-17 & Emergence of new white plumes WS2, 3 from the location of former barge B2 & Barge as the plume source. \\

2010-Nov-21 & Notable cloud holes and visibly-dark patches extending west along SEB(S) & \\

2010-Nov-19-30 & New plumes erupting from leading edge rather than original source. & \\

2010-Dec & Frequency and visibility of plumes begins to decrease. & Stored potential energy begins to be exhausted.\\

2010-Dec-11 & Northern branch reaches the GRS.  & \\

2011-Jan & Revival spread over $100^\circ$ longitude; first major dark spots on SEB(N). &  \\

2011-Jan-05 & Northern branch begins to form a dark collar around the GRS. & \\

2011-Jan-08 & Leading edge of Southern branch arrives at GRS. & SEB(S) revival almost complete. \\

2011-Feb-05 & Last reported plume from the source region. & End of convective activity. \\

2011-Feb-20 & Leading edge of central branch reached GRS. & \\

2011-Apr-06 & Solar Conjunction & \\

2011-May-Jun & Dark coloration restored at all longitudes. & \\

2011-Aug-Sep & General quiescence of SEB and formation of new, smaller cyclonic barges. & Potential for a renewed fade. \\

2011-Sep-21 & First signs of `normal' convection NW of the GRS. & Revival cycle has completed. \\

2011-Oct-09 & Jupiter at opposition & Centre of 2011/2012 apparition. \\

\hline
\end{tabular}
\end{center}
\label{tab:timeline}
\end{table*}%

\subsection{Initial SEB disturbance and revival source}
\label{seq0}

\begin{figure*}
\begin{centering}
\centerline{\includegraphics[angle=0,scale=0.9]{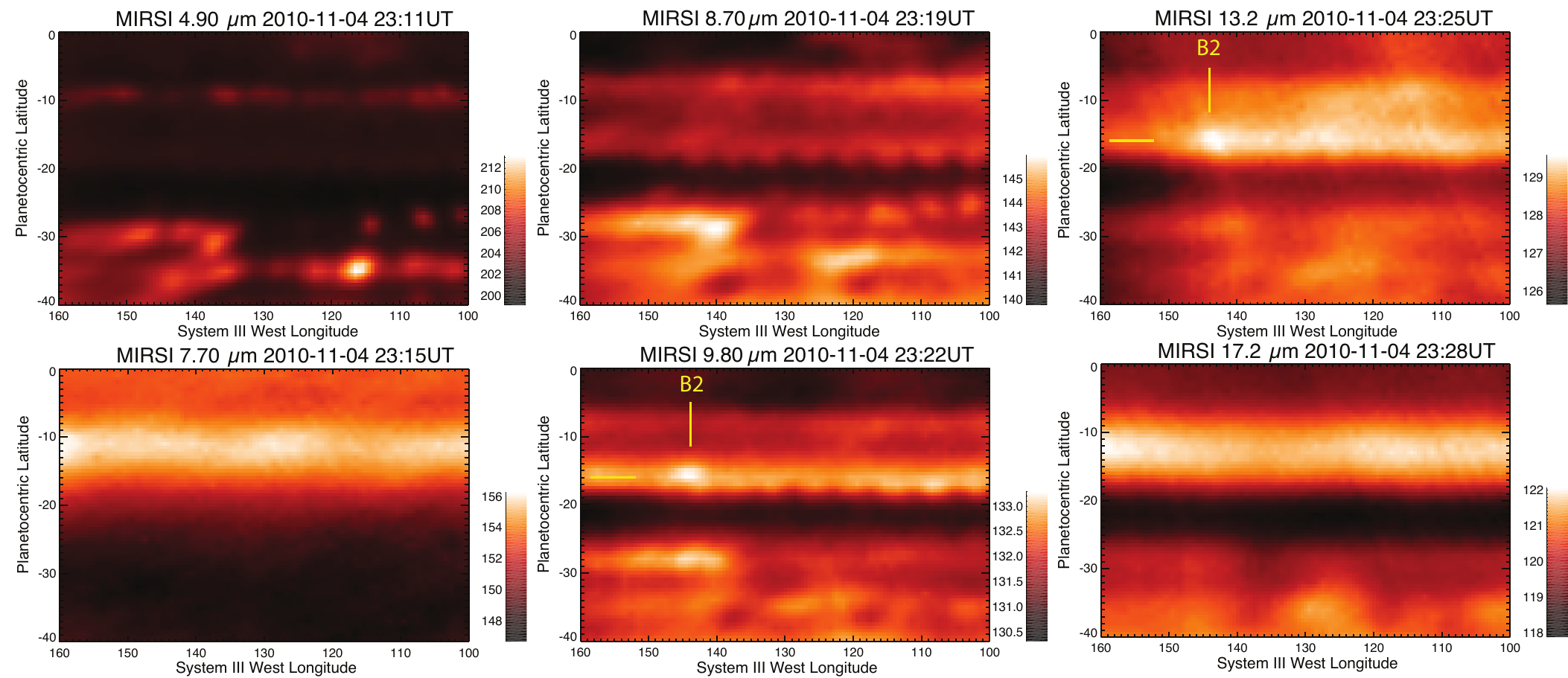}}
\caption{Survey of the faded SEB in IRTF/MIRSI imaging on November 4th 2010, five days before the outbreak at $148^\circ$W.  The 17.2 $\mu$m image is representative of similar images at 17.9, 18.4, 20.3 and 24.8 $\mu$m.  A warm spot is visible in the SEB near $144^\circ$W in the 9.8 and 13.2-$\mu$m filters (as indicated by the yellow lines), confirming the continued presence of a warm cyclonic barge immediately prior to the eruption.}
\label{prestorm}
\end{centering}
\end{figure*}

Amateur observers first noted a dramatic bright plume in the faded SEB on November 9th 2010 (day 0), which went on to become the source of the SEB revival \citep[white spot 1, WS1 at $15.1\pm0.2^\circ$S and $148^\circ$W,][]{11rogers_21, 16rogers}.  Prior to the eruption of WS1, IRTF/MIRSI infrared imaging on November 4th (Fig. \ref{prestorm}, five days before the outbreak) showed a relatively bland, quiescent SEB, with one notable exception.  Visible-light imaging had shown that WS1 erupted from the whitened remnant of cyclonic barge B2 that had been present earlier in 2010 \citep{11rogers_21, 12perezhoyos, 16rogers}.  The IRTF images in Fig. \ref{prestorm} confirm that this barge B2 remained warmer than its surroundings:  images at 9.8 $\mu$m (sensing a combination of $\sim400$-mbar temperatures and ammonia opacity) and 13.2 $\mu$m (sensing $\sim500$-mbar temperatures) show the presence of a small, warm sector at $16^\circ$S, $144^\circ$W, coincident with the whitened remnant of barge B2.  The sector spanned approximately $3^\circ$ longitude ($\sim3600$ km) and $2^\circ$ latitude. The cyclonic warm region was not observed at other wavelengths that sense higher, cooler altitudes, and did not produce cloud-free conditions sensed at 4.8 and 8.7 $\mu$m (Fig. \ref{prestorm}).

\begin{figure*}
\begin{centering}
\centerline{\includegraphics[angle=0,scale=.80]{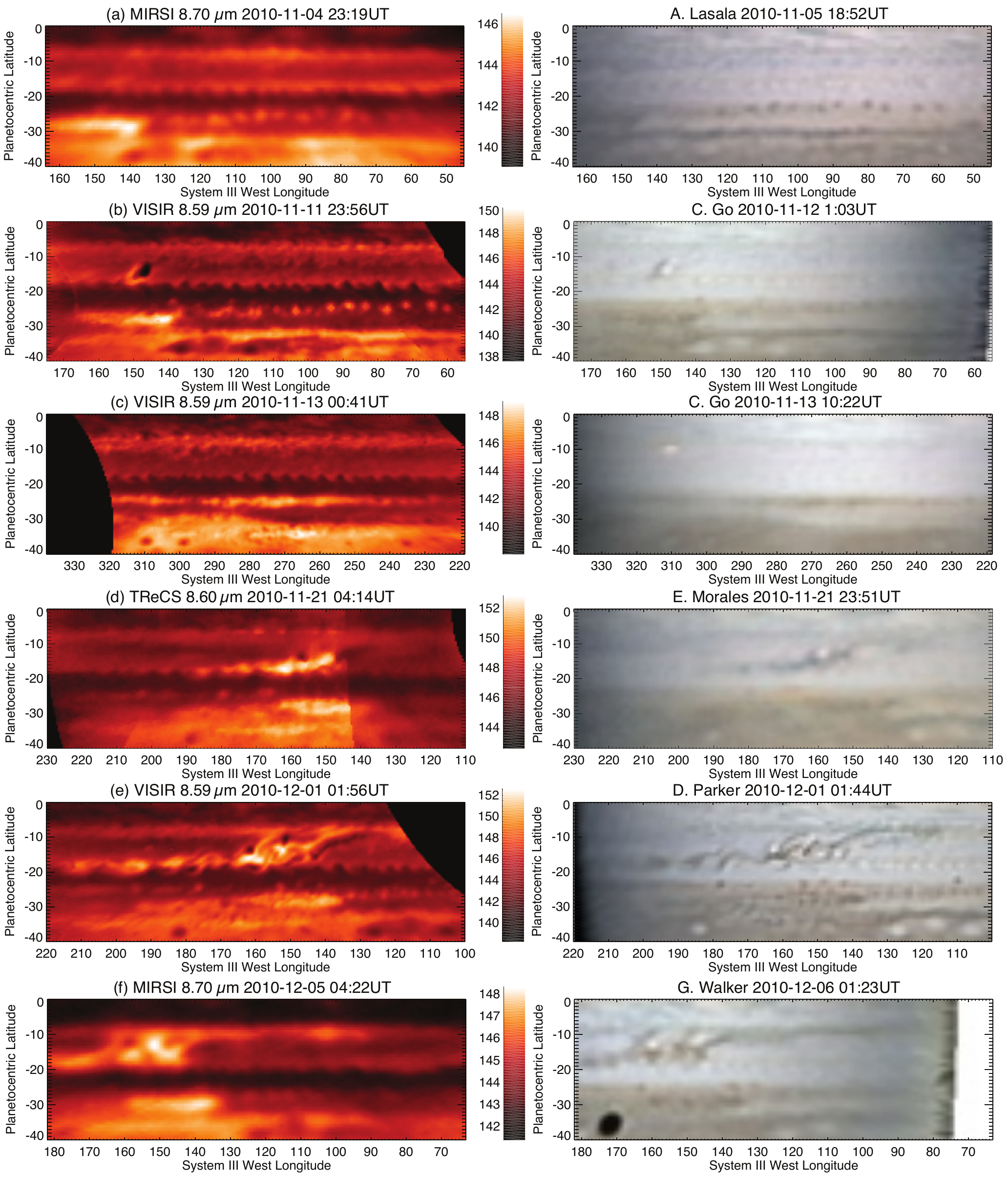}}
\caption{Comparison of 8.6-$\mu$m brightness temperature maps (sensing temperature and aerosol opacity near 600 mbar) and visible-light imaging from November-December 2010, showing the atmospheric conditions before the revival (a, c), during the initial outbreak (b), and following the development of plumes during the first month (d, e, f).  Dark oval-shaped regions in the thermal observations are due to an absence of data at these locations (either due to cropping by the instrumental field of view, or obscuration by chopped beams). The bright spot at $310^\circ$W in the visible image on row c is Io, the dark oval in the visible-light image in row f is a satellite shadow. The brightness temperature scale (K) is shown in the centre of the figure. }
\label{montage2010}
\end{centering}
\end{figure*}

After the eruption of WS1 from this cyclonic barge, the subsequent evolution of the plume is shown via a comparison of 8.6-$\mu$m and visible-light imaging in Fig. \ref{montage2010}.  Christopher Go (Philippines) first observed WS1 at 12:30UT on November 9th (Fig. \ref{montage2010}b), which was confirmed one rotation later by Donald Parker (USA) and Gary Walker (USA), when it was already brighter.  In subsequent rotations it was reported as the brightest spot on the planet by many observers, particularly in CH$_4$-band imaging at 889 nm, suggestive of a convective plume reaching high above the quiescent SEB cloud tops.  The first infrared image of this revival plume at $148^\circ$W was acquired by VLT/VISIR on November 11th (day 2, Fig. \ref{montage2010}b)\footnote{Circumstances outside of our control meant that only an 8.6-$\mu$m acquisition image was taken on November 11th rather than a full filter set}.  In addition, VISIR imaging was acquired two days later on November 13th (day 4), but targeted the still-faded SEB on the opposite hemisphere away from the erupting plumes (Fig. \ref{montage2010}c).  This confirms that there were no axisymmetric changes to the SEB once the plume eruption had started, and that the source of the revival was localised in longitude.  

Inspection of the revival plume (WS1) images on November 11th (day 2, Fig. \ref{montage2010}b) shows that the visibly-bright central plume was associated with a decrease in brightness temperature of $\sim10$ K compared to the surrounding faded SEB.  Furthermore, there was no obvious injection of latent heat from condensation at the altitudes probed by this filter.  Any warming associated with the first plume eruption must therefore be located at high pressures ($p>0.5$ bar), below the altitudes to which VISIR is sensitive.  The periphery of the plume appeared warmer and asymmetric, with a `tail' of brighter emission to the southwest that coincides with visibly-dark structures in the amateur imaging.  These dark structures usually formed on the westward side of the bright plumes within 2-3 days of the plume appearing \citep{11rogers_21, 16rogers}.  The structures subsequently extended northward to form a dark `lane' that marked the western boundary of a cell surrounding the plume.   The extension of the plume and peripheral warming to the southwest is consistent with the shear between the prograde SEBn jet near $7^\circ$S and the retrograde SEBs jet near $17^\circ$S.  

The decrease in brightness temperature associated with WS1 could be caused by two factors - adiabatic upwelling, expansion and cooling (near 500-600 mbar) within a rising plume, and/or a sharp rise in the aerosol optical depth to block out the 8.6-$\mu$m emission.  Without multi-spectral imaging we cannot distinguish between these two scenarios, but forward modelling indicates that the aerosol opacity must increase ten-fold at the plume location if temperatures are assumed to be constant.  Similarly, the slight increase in brightness temperature at the southwestern periphery is likely due to subsidence, adiabatic warming and the initiation of the ice sublimation to begin to reveal the dark underlying belt.  5-$\mu$m images obtained by IRTF/SpeX and IRTF/NSFCAM2 on November 13th and 16th, respectively (Fig. \ref{montage5um}a-b, day 4 to 7) do not readily show the individual plumes (which would appear dark at this wavelength due to their high opacity), but do reveal the increased emission from regions immediately west of WS1.  This decrease in aerosol opacity is consistent with subsidence through the 2-3 bar cloud deck, warming and clearing aerosols in the darker regions surrounding the plume.  WS1 was discernible for just five days \citep[the darker lane could be seen for eight,][]{11rogers_21, 16rogers}, but was just the first of many such eruptions at the revival source region.

\subsection{Plume Evolution in November 2010}
\label{seq1}

\begin{figure*}
\begin{centering}
\centerline{\includegraphics[angle=0,scale=.80]{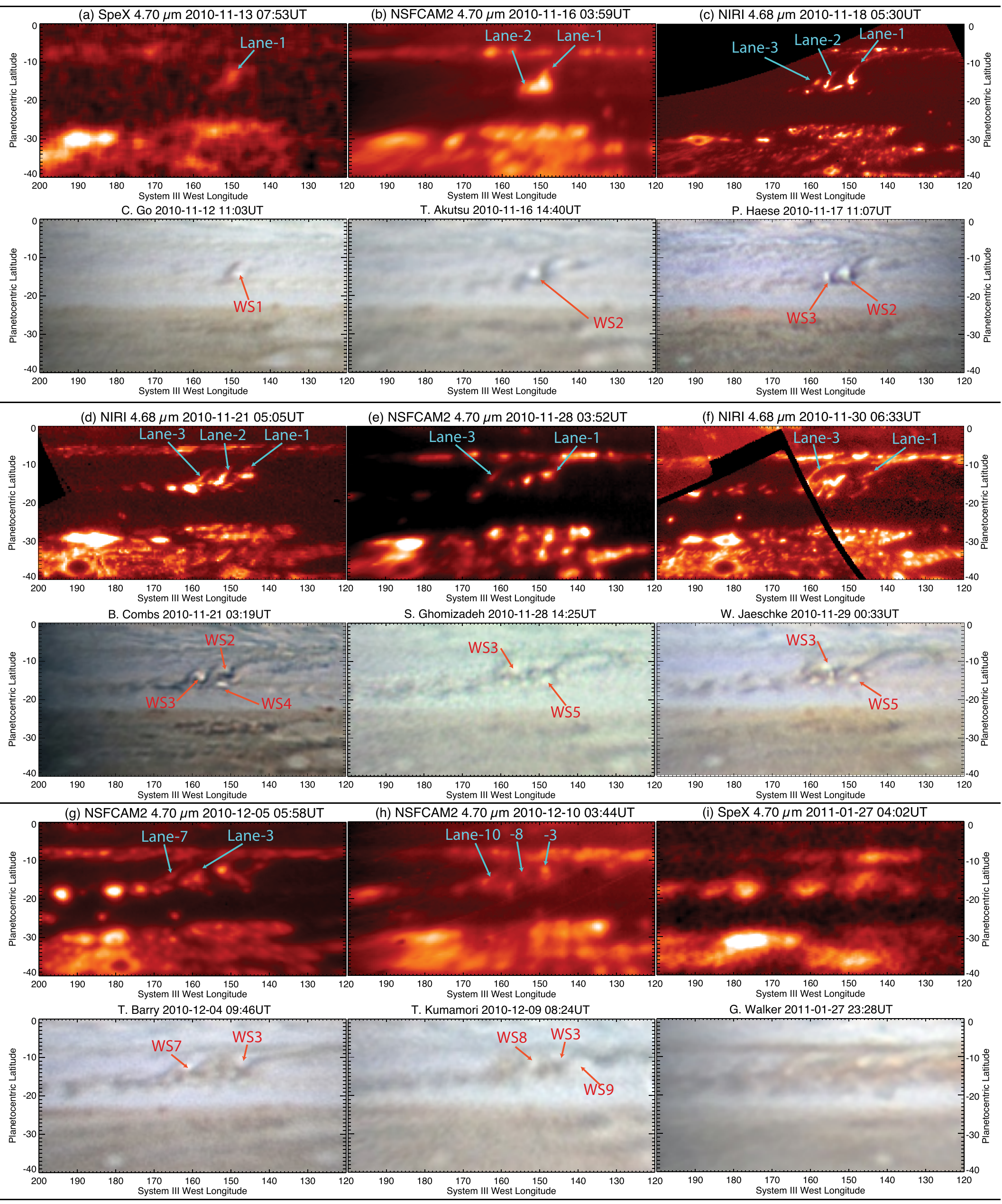}}
\caption{Comparison of 4.8-$\mu$m imaging and visible-light imaging from November 2010 to January 2011.  Each 4.8-$\mu$m image is accompanied by a visible-light image beneath it that was acquired nearby in time.  By combining observations from IRTF/SPEX, IRTF/NSFCAM2 and Gemini-N/NIRI (the latter of which has the highest spatial resolution), we are able to track the evolution of the central branch of the SEB revival centred on $160^\circ$W over the course of two months.  Note that the SPEX image on November 13th had to be heavily smoothed to improve the signal to noise, and that the NSFCAM2 images have a logarithmic scale to improve visibility of the small-scale contrasts.  The NIRI images on November 18th, 21st and 30th were formed from mosaics, so dark gaps can be seen where the images did not overlap.  These are uncalibrated counts, scaled to give the same contrasts in each panel.  The 4.8-$\mu$m maps are labelled with the locations of prominent bright lanes (dark in the visible images).  The visible maps are labelled with the location of white spots (WS), which sit at the centre of dark cells in the 4.8-$\mu$m maps. }
\label{montage5um}
\end{centering}
\end{figure*}

From November 9th to December 1st (days 0-22, Fig. \ref{montage2010} and Fig. \ref{montage5um}), \citet{11rogers_21,16rogers} reported three additional plumes erupting close to the original source of WS1, plus four similar plumes arranged in a line along the leading edge of the central branch (see below).   Each plume from the source subsequently appeared as the centre of a larger cell, bordered to the west by the darker lane.  Multiple plumes were sometimes active at the same time, and new plumes appeared every 2-6 days between November 9th and December 13th 2010.  By December 2010, each cell was developing a reddish-brown coloration, particularly in their southern parts.  Together, these cells comprised the central branch of the SEB revival. 

Between November 18th and 30th (day 9 to day 21), 5-$\mu$m imaging from Gemini-N/NIRI studied the deep cloud structure in the central branch on three occasions, using lucky imaging to provide exceptional spatial resolution (Fig. \ref{montage5um}c, d, f).  The brightest emission on November 18th (Fig. \ref{montage5um}c) occurred as `comma-like' lanes reaching from the SEB(N) in the northeast to SEB(S) in the southwest, coinciding with the dark lanes in visible light.  These structures were associated with WS2 (November 14th-20th) and WS3 (Nov 17th - Dec 13th) tracked at the source region by \citet{11rogers_21}.  By November 21st (day 12, Fig. \ref{montage5um}d), the bright patches of 5-$\mu$m emission extended to the west along the SEB(S) - these cloud holes formed the southern branch of the disturbance, and coincided with darker patches travelling along the retrograde SEBs jet.   Between November 19th and 28th, \citet{11rogers_21} reports that new plumes (WS4, 5 and 6) were forming not at the original source, but in a line on the leading edge of the central branch located a few degrees to the east.  A visibly-dark lane was associated with this edge, which can be seen as a 5-$\mu$m-bright `zig-zag' lane stretching across the SEB near $147^\circ$W on November 30th (day 21, Fig. \ref{montage5um}f). 

Fig. \ref{montage5um} reveals at least three interrelated phenomena at work - (i) the bright white plumes in visible light erupting both at the original source and at the leading edge of the disturbance; (ii) the strong cloud-clearing and bright 5-$\mu$m emission immediately adjacent to the plumes; and (iii) the moderate cloud-clearing associated with the narrow lanes crossing the SEB.  As time progressed it became harder to distinguish these regions as more plumes and lanes were formed.  

The November 21st Gemini-N/NIRI 5-$\mu$m observation (Fig. \ref{montage5um}d) was obtained within an hour of a Gemini-S/TReCS 8.6-$\mu$m observation (Fig. \ref{montage2010}d).  Whereas the 5-$\mu$m emission (sensing holes in the 2-3 bar clouds) is highly localised, the 8.6-$\mu$m emission (sensing thermal contrasts and cloud holes at ~600 mbar) appears more diffuse, with the cloud-free bright regions blending into a more continuous band than can be seen at 5 $\mu$m.  This hints at complex vertical structure in the downwelling branch of the convective patterns that affects the aerosol fields in different ways at different altitudes - 5-20 $\mu$m spectroscopy would have been highly desirable to understand this vertical aerosol structure.

\subsection{Plume complex in December 2010}
\label{seq2}

\begin{figure*}
\begin{centering}
\centerline{\includegraphics[angle=0,scale=.90]{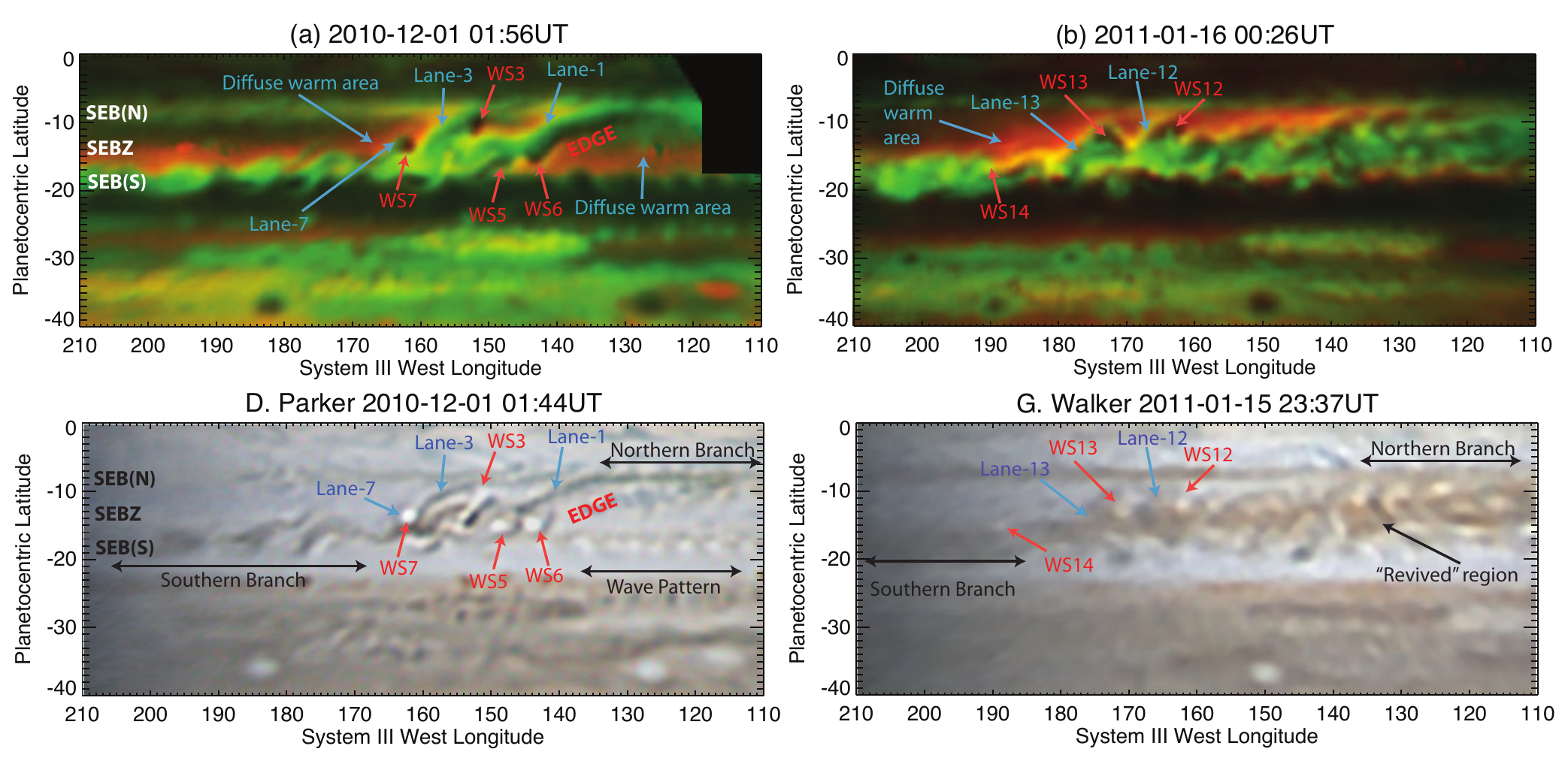}}
\caption{Labelled images of the SEB revival region on (a) December 1st 2010 and (b) January 16th 2011.  Thermal-IR images from VLT/VISIR are shown as red-green (RG) composites, using 8.6 $\mu$m as the green channel (sensing aerosol opacity) and 10.8 $\mu$m as the red channel (sensing 500-mbar temperatures and ammonia).  Near-simultaneous visible-light images are shown for comparison from D. Parker and G. Walker.  Prominent SEB features - the plumes (WS) and associated lanes, the northern and southern branches, are all labelled, as well as the wave pattern identified by \citet{16rogers_wave}.  WS3 and 7 erupted at the source, WS5 and 6 erupted on the leading edge of the disturbance (the remnant of WS1).  Warm and cloudy regions appear red, cold and cloud-free regions appear green, cold and cloudy regions appear black, and warm and cloud-free regions appear yellow.}
\label{visir_label}
\end{centering}
\end{figure*}

\begin{figure*}
\begin{centering}
\centerline{\includegraphics[angle=0,scale=.80]{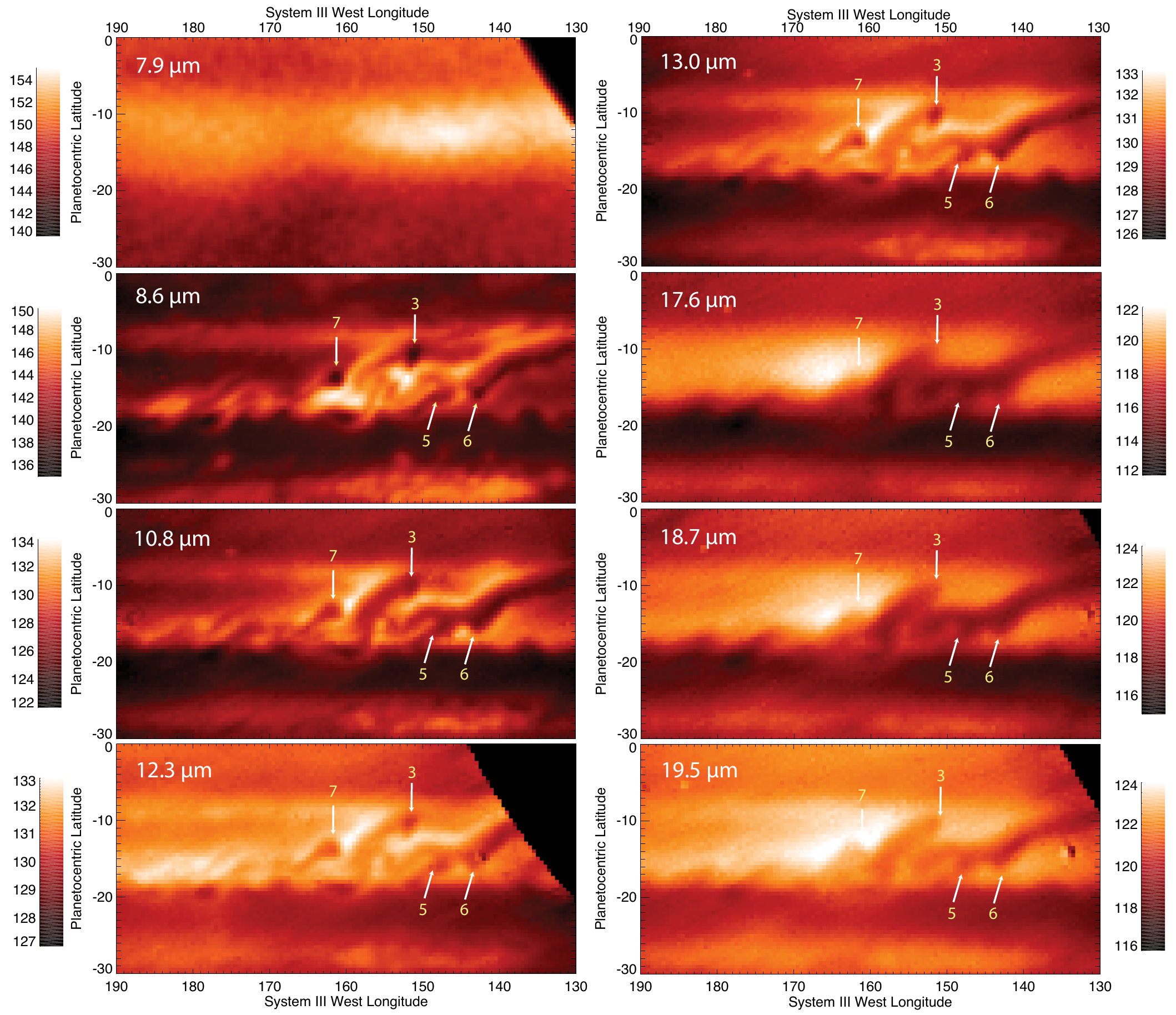}}
\caption{Brightness temperatures from December 1st 2010 (01:52-02:31UT) shown in all eight VISIR filters.  The brightness temperature scales are given in Kelvin.  A labelled RG composite of the 8.6- and 10.8-$\mu$m filters is compared to visible-light imaging in Fig. \ref{visir_label}, but we point out the locations of four white plumes (arrows) in each panel.}
\label{filters}
\end{centering}
\end{figure*}

Our first multi-spectral view of the revival complex was obtained by VLT/VISIR on December 1st 2010 (day 22), by which time more plumes had appeared and a significant section of the SEB ($140-160^\circ$W) had been warmed by the disturbance (Fig. \ref{montage2010}e).  Fig. \ref{visir_label}a provides a labelled map of the revival region on this date.  The two-colour RG composite image uses brightness temperatures at 8.6 $\mu$m as the green channel (where higher brightness suggest low aerosol opacities) and 10.8 $\mu$m as the red channel (where higher brightness suggests a warmer atmosphere or low NH$_3$ gas opacity, Table \ref{tab:filters}).  In addition, Fig. \ref{filters} shows the appearance of the central branch in all eight filters between 7.9-19.5 $\mu$m, with the sensitivity to temperature and composition detailed in Table \ref{tab:filters}.  Images in the right hand column (13-20 $\mu$m) sense atmospheric temperatures via the H$_2$-He collision-induced continuum, whereas images in the left hand column sense combinations of temperature, ammonia humidity, aerosol opacity and stratospheric hydrocarbon emission (e.g., the 7.9-$\mu$m image senses the stratospheric temperatures).    

The images in Figs. \ref{visir_label}a and \ref{filters} reveal the central branch, the leading edge of the disturbance to the southeast, and reviving southern branch extending to the west (the SEB(S)).  These features combine to give the revival an `S-shaped' appearance, being sheared eastward by the prograde SEBn jet and westward by the retrograde SEBs jet.   Using the nomenclature of \citet{16rogers}, we identify three major features in the December-1st images:  (a) the cold plume of WS3 at $-10.5^\circ$S, $152^\circ$W, visible from November 17th to December 13th near to the initial location of WS1, with its associated dark lane to the west.  (b) WS7 near $-14^\circ$S, $162^\circ$W, first seen on this date and persisting until December 9th.  (c) The substantial dark lane between $130-145^\circ$W that had been associated with WS1, which had itself disappeared on November 14th.  The size of WS3 and WS7 can be crudely estimated from the 8.6-13 $\mu$m images as circles with $1.5^\circ$ diameter, equivalent to 1800 km (assuming one degree of longitude represents 1215 km at $14^\circ$S) or an area of $2.5\times10^6$ km$^2$ (approximately equivalent to the area of the Mediterranean Sea).  In addition to these plumes from the source, the leading edge of the disturbance shows temperature contrasts associated with additional plumes \citep[WS5 and WS6,][]{11rogers_21}.  These bright plumes appeared concentrated along this line and separate from those in the source region, and several appeared between December and January at a range of latitudes along this leading edge.  

The RG composite in Fig. \ref{visir_label}a shows the plumes and the leading edge to be cold and cloudy (appearing black).  Regions between the eastern edge at $\sim140^\circ$W and the source near $\sim160^\circ$W appear cold but clear of cloud opacity (showing up as green), showing the advance of the revival to the east of the source region.  Regions to the northwest of the central branch and to the southeast of the leading edge both appear warm and cloudy (appearing as diffuse red).   These diffuse regions appear warm not only at 10.8 $\mu$m (which is sensitive to ammonia opacity), but also at 13.0-19.5 $\mu$m (which are sensitive only to temperature). The atmosphere is warmest to the west of the plume source region ($160-170^\circ$W), and this can be readily seen in the Q-band images (17.6-19.5 $\mu$m in Fig. \ref{filters}).  The comparison in Fig. \ref{visir_label} shows that this diffuse warm glow is associated with the whitened SEB, whereas low albedo features (i.e., reviving colours) are instead associated with regions of high 8.6-$\mu$m flux revealing gaps in the clouds (green and yellow colours in the RG composite).  This suggests that the diffuse temperature rise is occurring in the 100-400 mbar region \citep[within the upper tropospheric hazes of][]{12perezhoyos}, and has not yet propagated to sufficient depth to modify and clear the white aerosol opacity of the ammonia-cloud deck ($1.0\pm0.4$ bar).

\subsection{Central, Northern and Southern Branches - January and February 2011}
\label{seq3}

\begin{figure*}
\begin{centering}
\centerline{\includegraphics[angle=0,scale=.80]{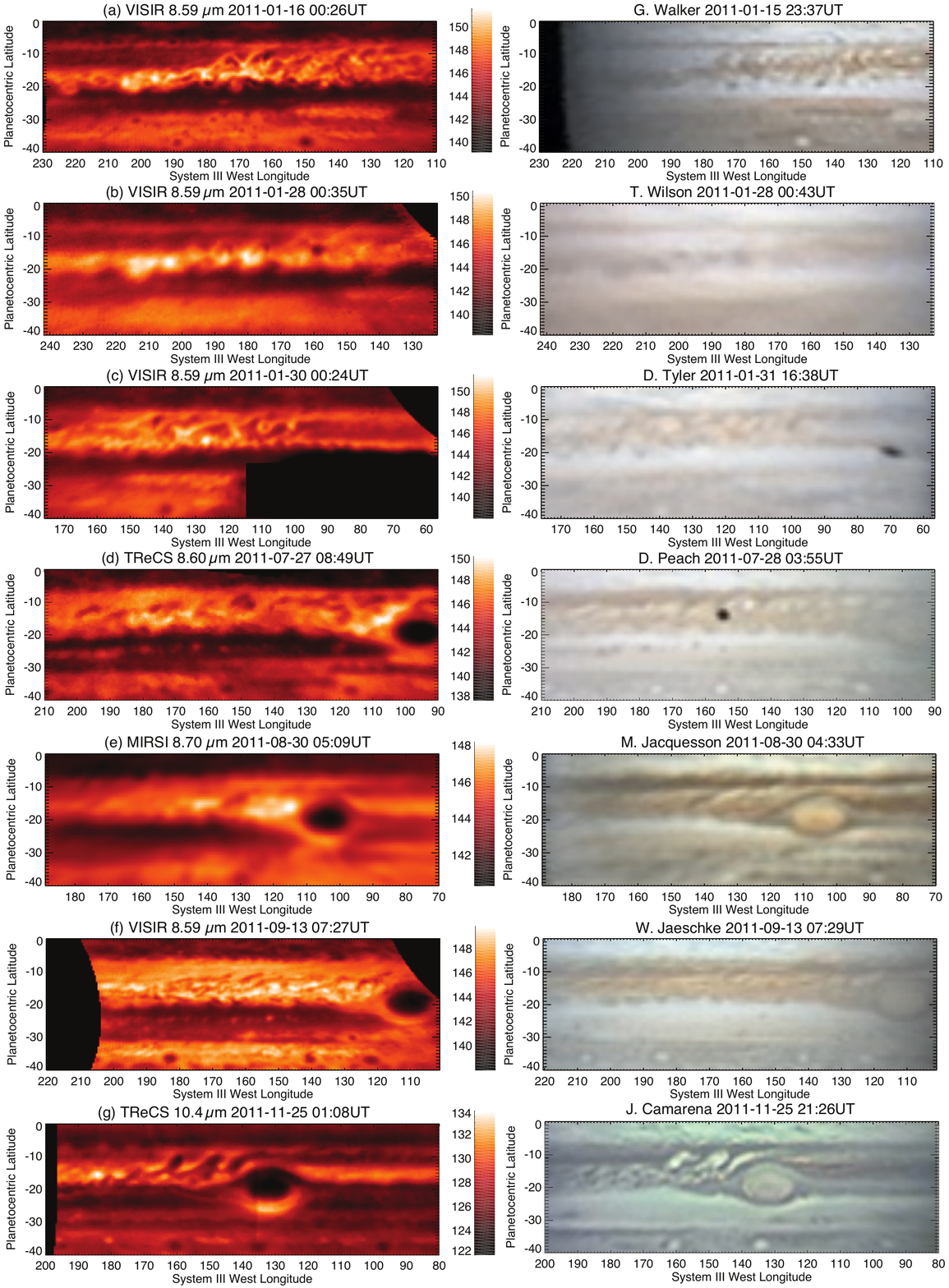}}
\caption{Comparison of 8.6-$\mu$m brightness temperature maps and visible-light imaging from January-September 2011, with an additional 10.4-$\mu$m Gemini-S/TReCS image in November 2011 after the GRS `rifting' had been re-established (g).  VISIR data in January and September 2011, along with COMICS data in August 2011, are assembled into full cylindrical maps in Figs. \ref{cmap2011jan}, \ref{cmap2011aug} and \ref{cmap2011sep}.  Dark shadows in the visible-light images in rows (c) and (d) are due to satellite shadows. The brightness temperature scale (K) is shown in the centre of the figure.  The black rectangle to the lower right of panel (c) masks a chop-obscured region of Jupiter's disc.}
\label{montage2011}
\end{centering}
\end{figure*}

Figs. \ref{visir_label}b and \ref{montage2011} show the later stages of the revival from January 16th 2011 onwards (day 68), by which time the typical brown appearance of the SEB was becoming apparent over the $100-200^\circ$W longitude range. \citet{11rogers_21} reports that the frequency and visibility of the plumes decreased after mid-December, and the pattern of `comma-shaped' cells was no longer apparent by January.  The lifetime of the individual plumes was 4-22 days and 15 such plumes were observed between November and early January \citep{11rogers_21, 16rogers}.  Individual plumes continued to erupt form the source region until February 5th (the final white spot observed, although few observations were possible thereafter before solar conjunction began), but the whole revival had become too complex to track individual spots.  This is evident from the RG composite on January 16th in Fig. \ref{visir_label}b, where multiple cold plumes (WS12, 13 and 14 are labelled) are embedded within a complex network of bright filaments that represent the widening gaps in the 700-mbar aerosol layer.  This image clearly shows the correspondence between the brown colours and enhanced 8.6-$\mu$m emission, along with peripheral warming within the narrow, visibly-dark lanes and in the diffuse red glow to the northwest of the central branch. 

\begin{figure*}
\begin{centering}
\centerline{\includegraphics[angle=0,scale=.85]{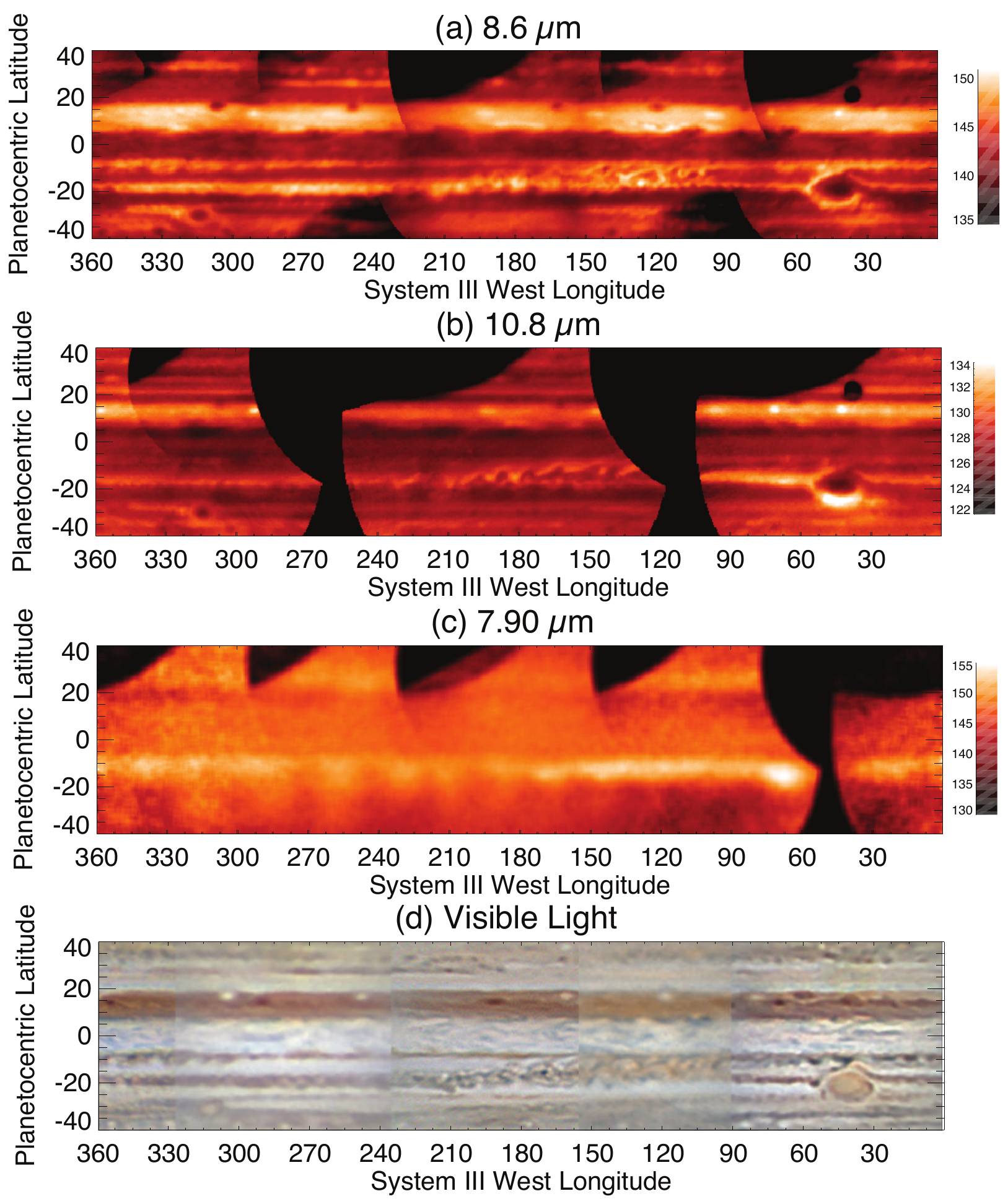}}
\caption{Near-global brightness-temperature maps of Jupiter at 8.6, 10.8 and 7.9 $\mu$m acquired by VLT/VISIR between January 27 and 31, 2011, during the mature phase of the SEB revival.  Black segments of the planet indicate locations not observed by VISIR's field of view.  This is compared to a visible light map in panel d assembled by M. Vedovato from images between January 28 and February 1st by G. Jolly, G. Walker, D. Tyler, and D. Parker.  The brightness temperature scale (K) is shown to the right of each map.}
\label{cmap2011jan}
\end{centering}
\end{figure*} 

\textbf{Central branch: } Our final thermal-IR observations of the 2010-11 apparition were acquired by VLT/VISIR between January 27th-31st 2011 (day 79 to day 83).  By combining images over four nights, we assembled a near global map of Jupiter during the revival (Fig. \ref{cmap2011jan}), the first of its kind from an 8-m diameter telescope.  This shows that the turbulent activity within the central branch dominated longitudes between 100-180$^\circ$W, but with extensions to the east and west along the northern and southern branches.  The central branch appeared cloud-free (bright at 8.6 $\mu$m) and orange-brown in colour.  The 10.8-$\mu$m image, sensing a combination of temperatures and ammonia humidity near 500 mbar, shows a distinct meandering lane of warmth extending from $210^\circ$W on the SEB(S) all the way to $120^\circ$W on the SEB(N).  To the northwest of this lane we see diffuse warm temperatures in regions that remain white in colour (i.e., sublimation has yet to take place), to the south east we see cooler temperatures, presumably associated with upwelling and adiabatic expansion in the remaining plumes.  Intriguingly, the thermal contrasts at 10.8 $\mu$m appear less complex than the chaotic cloud contrasts sounded at 8.6 $\mu$m.  The central branch is clearly warming the SEB compared to regions unaffected by the revival, but the plumes themselves still appear to be cold.  

\textbf{Northern branch: } Besides the central branch, Fig. \ref{cmap2011jan} shows that the northern branch of the SEB revival comprises a narrow band of dark material in the visible images, although changes to this band didn't begin until mid-December.  \citet{11rogers_21} reports a dark segment of the SEB(N) moving east away from the central branch, reaching the GRS on December 11th and continuing onwards.  The northern branch became more vigorous after January 5th, comprising a chain of dark spots that moved past the northern edge of the GRS \citep{11rogers_24}.  These dark undulations on the SEB(N) can be seen in Fig. \ref{cmap2011jan}(d), just north of the GRS.  The interaction of the northern branch with the Red Spot Hollow began around January 5th (day 57), beginning to form a dark collar around the GRS in visible light images.  Although this longitude was not observed at 5 $\mu$m in January 2011, we speculate that this dark collar corresponds to a clearing of aerosol opacity surrounding the GRS, allowing the bright peripheral ring to reform - it was easily visible by August 2011 in Fig. \ref{5um_maps}(c).

\textbf{Southern branch: } The southern branch of the revival was more complex.  The SEB(S) had almost completely disappeared at 8.6 $\mu$m prior to the revival (i.e., it was completely clouded over), but Fig. \ref{cmap2011jan} shows multiple sectors of bright 8.6-$\mu$m emission that correlate with visibly-dark westward-moving patches in amateur imaging.  These patches are likely regions of newly-cleared atmosphere that are moving westward along the retrograde jet.  However, not all dark spots were bright in the infrared (Fig. \ref{5um_maps}), suggesting some inhomogeneity in the cloud-clearing processes occurring in the southern branch.  We cannot rule out the possibility that some of the visibly-dark regions are associated with the production of dark, soot-like material \citep{09baines_storms}.  

By late-January, the SEB(S) was fully revived all the way from the source region to the Great Red Spot, the westward-moving southern branch having arrived there around January 8th.  The sector immediately west of the GRS, between the leading edge of the central branch and the GRS, appears warmer at 10.8 $\mu$m compared to the rest of the SEB(S), and yet the dark albedo and 8.6-$\mu$m emission is at its most latitudinally-confined over this longitude range ($50-100^\circ$W).  The leading edge of the central branch finally reached the GRS on February 20th (day 103), such that the entire SEB from $180^\circ$W to the GRS at $50^\circ$W was revived \citep{11rogers_24, 16rogers}.  By the end of February, Jupiter was heading for solar conjunction (April 6th 2011) and was unavailable to ground-based visible or infrared observatories. 

\subsection{Stratospheric Effects of the SEB Revival}
\label{stratwave}
Paper 1 concluded that the fade of the SEB had no consequences for stratospheric temperatures.  Before the emergence of the revival plume, IRTF/MIRSI imaging (Fig. \ref{prestorm}) showed the presence of a longitudinally-uniform warm band in the $10-15^\circ$S region.  VLT/VISIR observed Jupiter's CH$_4$ emission during the first 100 days of the revival, on December 1st, January 16th and January 27th-31st.  Fig. \ref{filters} shows the first observation, with notably warmer temperatures between $140-160^\circ$W (i.e., directly above the source region) compared with other longitudes.  The near-complete map in late January 2011 (Fig. \ref{cmap2011jan}(c)) reveals that this was the start of a stratospheric wave pattern that had not been present prior to the revival (it was not observed in VISIR images on November 13th for longitudes away from the revival source).   The wave pattern is at its most regular in the January 2011 images - maps from the next apparition (in August and September 2011, Figs. \ref{cmap2011aug} and \ref{cmap2011sep}, discussed below) show longitudinal stratospheric variability above the SEB, but not with the same regularity.

Fig. \ref{cmap2011jan}(c) therefore reveals, for the first time, a stratospheric response to the powerful convective activity involved in the central branch of the revival.  The wave is clearly visible to the west of the central branch, from $150-330^\circ$W, covering a broader longitude range than the central branch itself.  Indeed, the longitudinal coverage more closely resembles the range covered by the newly-revived SEB(S) in January 2011.  Eight or nine distinct maxima can be observed, resulting in a wavelength of $20-30^\circ$ longitude.  The central latitude of the wave peaks is $10-15^\circ$S, but the waves appear to show southward extensions to at least $30-40^\circ$S.  These elongated structures suggest that the SEB revival is affecting stratospheric temperatures not just over the SEB itself, but over a broader latitude range encompassing Jupiter's tropical and temperate domains.  This elongation was only observed once, in January 2011, but was a consistent feature across multiple independent images over the four nights of observation.  The wave pattern was not visible in reflected light in the methane absorption band at 0.89 $\mu$m (amateur images from January 11th to February 8th; data not shown).

Unfortunately this snapshot of wave activity is insufficient to determine the nature of the wave.  We speculate that the powerful convective plumes served as a wave source (gravity and Rossby waves), which were able to propagate vertically and manifest as this undulating thermal pattern in Jupiter's stratosphere.  A similar connection between stratospheric waves and tropospheric meteorology was observed during Saturn's storm \citep{12fletcher}, and Section \ref{discuss} provides a detailed comparison of these two planetary-scale events.

%

\subsection{Completion of the Revival - July-September 2011}
\label{seq4}

The start of the Jupiter apparition in May-June 2011 (day $\sim200$) allowed amateur observers to track the progress of the SEB revival, as summarised by \citet{15rogers, 16rogers}.  Fig. \ref{5um_maps}c-d shows that the normal dark coloration and 5-$\mu$m emission had been restored at all longitudes by the start of the apparition, with some final convective white spots emerging from the revival outbreak region to the west of the GRS \citep{15rogers}.  These spots moved eastward towards the GRS without any new outbreaks in August, and by September (day 300) very little evidence of the original outbreak was left.  The GRS was pale orange in colour and surrounded by a dark grey peripheral ring that was devoid of cloud opacity, appearing bright at 5 $\mu$m (Fig. \ref{5um_maps}c).  The turbulent wake northwest of the GRS continued to be quiescent during this period, with a distinct reddish tinge in images between May and September.  On September 21st (day 316), the first convective white spot appeared in the GRS wake since the start of the revival, initiating the normal convective activity that has continued since that time.  As time progressed, these plumes became increasingly visible in methane-band imaging sensing the upper troposphere, and signalled the return of the SEB to its normal conditions.

\begin{figure*}
\begin{centering}
\centerline{\includegraphics[angle=0,scale=.85]{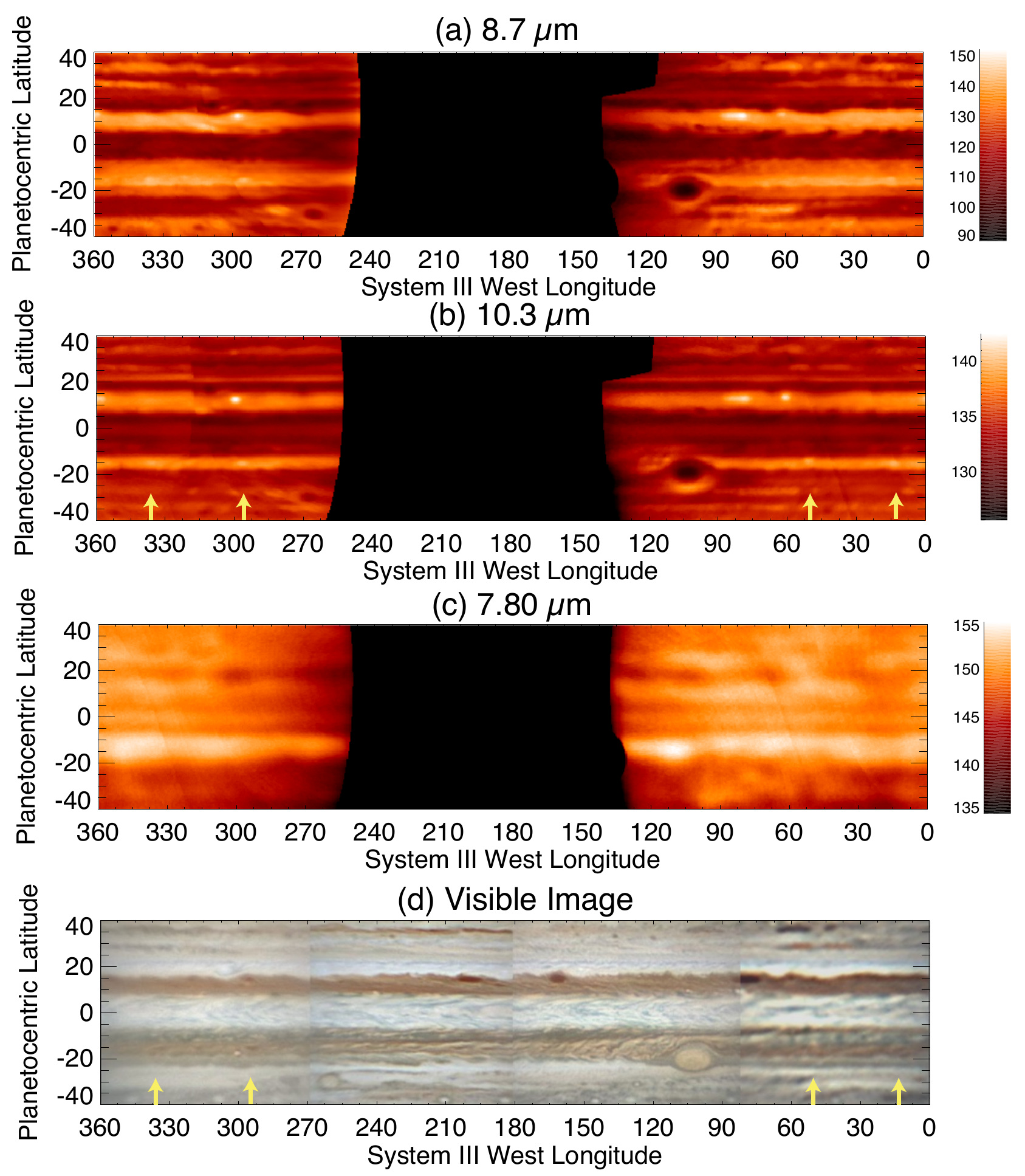}}
\caption{Brightness temperature maps of Jupiter at 8.7, 10.3 and 7.8 $\mu$m acquired by Subaru/COMICS on August 27th 2011 between 13:00-15:40 UT, during the mature phase of the SEB revival.  Black segments of the planet indicate locations not observed by COMICS's coverage. This is compared to a visible light map in panel d assembled by M. Vedovato from images between August 29th-30th by M. Jacquesson, C. Go and A. Wesley.  The brightness temperature scale (K) is shown to the right of each map.  The location of newly-formed barges is shown by the yellow arrows.}
\label{cmap2011aug}
\end{centering}
\end{figure*}

\begin{figure*}
\begin{centering}
\centerline{\includegraphics[angle=0,scale=.85]{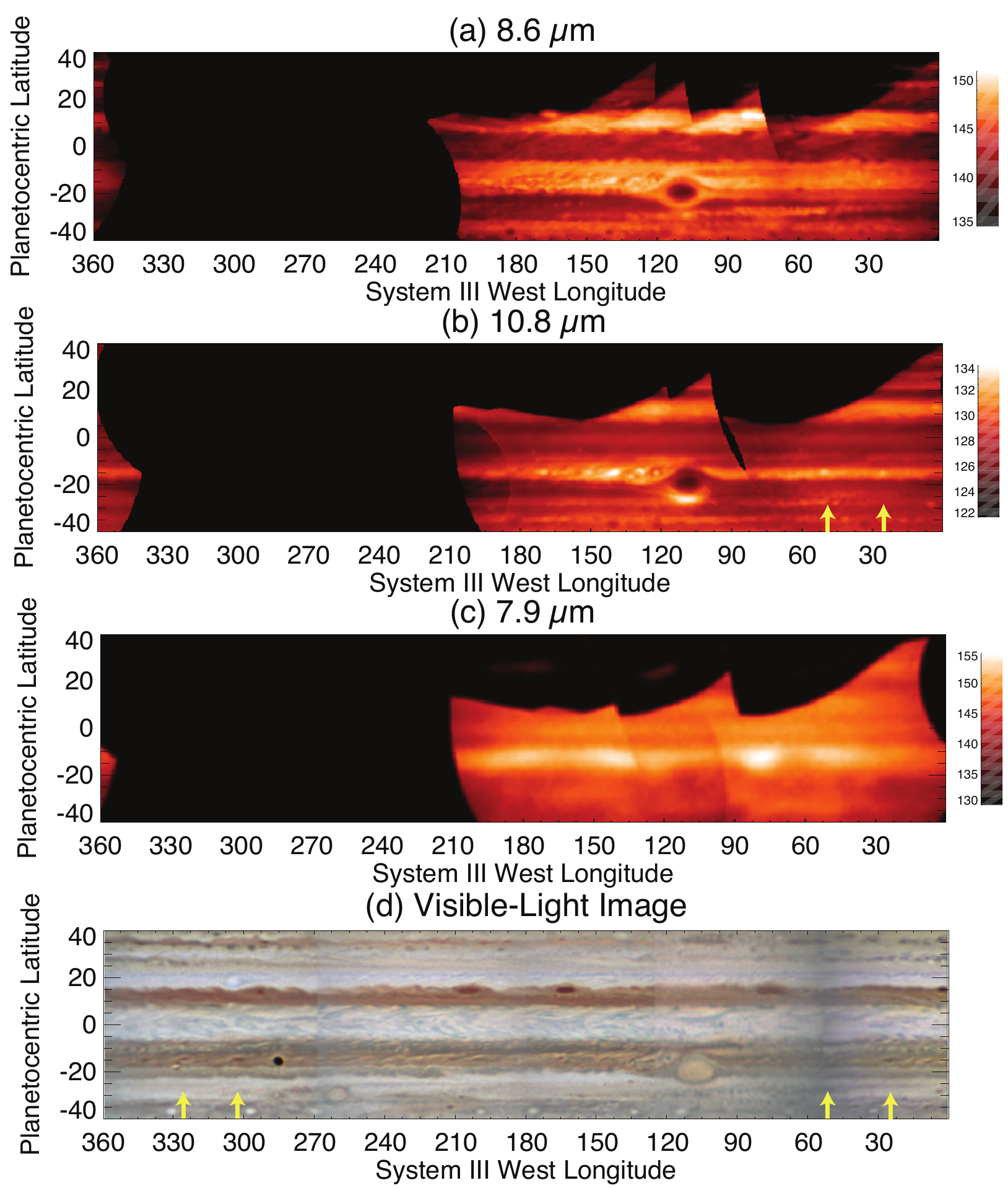}}
\caption{Near-global brightness temperature maps of Jupiter at 8.6, 10.8 and 7.9 $\mu$m acquired by VLT/VISIR between September 13-18th 2011, during the last phase of the SEB revival.  Black segments of the planet indicate locations not observed by VISIR's field of view.  This is compared to a visible light map in panel d assembled by M. Vedovato from images between September 18th and 19th by I. Sharp, B. Combs, E. Morales and D. Peach. The brightness temperature scale (K) is shown to the right of each map. The location of newly-formed barges is shown by the yellow arrows. }
\label{cmap2011sep}
\end{centering}
\end{figure*}

Thermal infrared imaging of this mature phase, before the new GRS rifting began, was captured by Gemini-S/TReCS (July 27th, Fig. \ref{montage2011}d, day 260), Subaru/COMICS (August 27th, Fig. \ref{cmap2011aug}, day 291), IRTF/MIRSI (August 30th, Fig. \ref{montage2011}e, day 294) and VLT/VISIR (September 13th-18th, Fig. \ref{montage2011}f and Fig. \ref{cmap2011sep}, day 308-313).  The TReCS 8.6-$\mu$m image on July 27th confirms the now-familiar pattern: brighter albedo regions in the SEB clouds show as dark and cold (i.e., more opaque) at 8.6 $\mu$m, whereas darker structures appear bright and aerosol-free. 

\textbf{The warm SEB(S):} Although the Subaru/COMICS maps in August 2011 missed the original source region, the comparison of the 8.6-$\mu$m and 10.3-$\mu$m image in Fig. \ref{cmap2011aug} reveals a broad warm 10.3-$\mu$m band over the SEB(S) near $15^\circ$S that coincides with the darkest brown coloration, and a second fainter warm band at the SEB(N).  These bands can also be observed in the VLT/VISIR imaging in similar wavelengths in September 2011 (Fig. \ref{cmap2011sep}). The narrow warm bands did not coincide with the broader aerosol-free region observed at 8.6 $\mu$m, which extended over the full latitudinal width of the SEB.  Between the two warm bands lies the cooler SEB zone (SEBZ), a characteristic of the SEB for longitudes away from the GRS.  The southern warm band was not present in January 2011 in the still-faded regions of the SEB (Fig. \ref{cmap2011jan}(b)).  However, it was present with less contrast during the fade in November 2009 and July 2010 (Figs. 4b, 5b and 5c of Paper 1), and in the November 13th 2010 images presented in this paper (Fig. \ref{Tmap}, see below).  The warm SEB(S) is likely to be the `normal' state of this band for longitudes far from the GRS, and it was only obscured by elevated aerosol opacity during the peak of the fade.

\textbf{Post-revival quiescence: } Embedded within the 10.3-$\mu$m bright band are small oval-shaped regions of even warmer conditions (yellow arrows in both Fig. \ref{cmap2011aug} and \ref{cmap2011sep}).  These small warm patches are at the spatial-resolution limit of the visible and thermal-IR imaging, but appeared to be smaller versions of the five deep-red, warm cyclonic barges that had been present in 2010 during the fade (Paper 1).  Approximately four of these warm barges could be seen between $300-360^\circ$W and $0-60^\circ$W in August and September 2011.  Unlike the more substantial barges observed during the fade (see Fig. 2 of Paper 1), these smaller barges could not be distinguished at 8.6 $\mu$m.  The formation these barges, coupled with the general quiescence of the SEB, had led to speculation that a new fade might begin \citep{15rogers, 16rogers}.  This post-revival quiescence has been noted in previous cycles \citep[1971, 1975, 1990, 1993,][]{16rogers}, and a renewed fade did start after both the 1975 and 1990 revivals.  However, on this occasion the renewal of convective activity west of the GRS in late September 2011 meant that the revival was complete, with no new fading cycles evident between September 2011 and December 2016 (the time of writing).  

Gemini-S/TReCS offered a final 10.4-$\mu$m observation in Fig. \ref{montage2011}g, sensing tropospheric temperatures and ammonia humidity, on November 25th 2011, twelve months after the revival outbreak.  Although this single acquisition image was not part of a full multi-wavelength set, it reveals the cold filamentary rifts and adjacent warm lanes associated with the convective activity in the GRS wake.  The pattern of this rifting is identical to that seen in January 2011 during the revival (Fig. \ref{cmap2011jan}b), suggesting that the same processes - narrow, cold, convective plumes surrounded by regions warmed by subsidence - are at work in both the GRS wake and the SEB revival.  These processes will be explored in Section \ref{discuss}.

\subsection{Temperature and aerosol changes during the revival}
\label{summary}

Using identical techniques to Paper 1, we employed the NEMESIS retrieval algorithm to invert the crude spectra formed by stacking the 7-20 $\mu$m narrow-filter imaging from VLT/VISIR \citep{08irwin, 09fletcher_imaging}.  This allows us to map thermal and aerosol contrasts at key stages during the revival in Fig. \ref{Tmap}, from the faded state (November 2010) to the fully-revived state (September 2011).  Tropospheric temperatures in the 200-700 mbar range are derived from H$_2$-He-sensitive filters at 13.0, 17.6, 18.7 and 19.5 $\mu$m (along with 8.6 and 10.8 $\mu$m observations that also sense aerosols and ammonia), whereas stratospheric temperatures near 5 mbar are derived from CH$_4$ emission near 7.9 $\mu$m and, to a lesser extent, C$_2$H$_6$ emission near 12.3 $\mu$m (Table \ref{tab:filters}).  With only eight filtered points defining the spectrum, formal uncertainties in the derived temperatures are large (ranging from $\sim3$ K in the troposphere to $\sim5$ K in the stratosphere). However, these uncertainties in the absolute temperatures are much larger than the pixel-to-pixel relative uncertainties, which have a precision of 0.5-1.0 K.

\subsubsection{Temperature variations}
\begin{figure*}
\begin{centering}
\centerline{\includegraphics[angle=0,scale=.95]{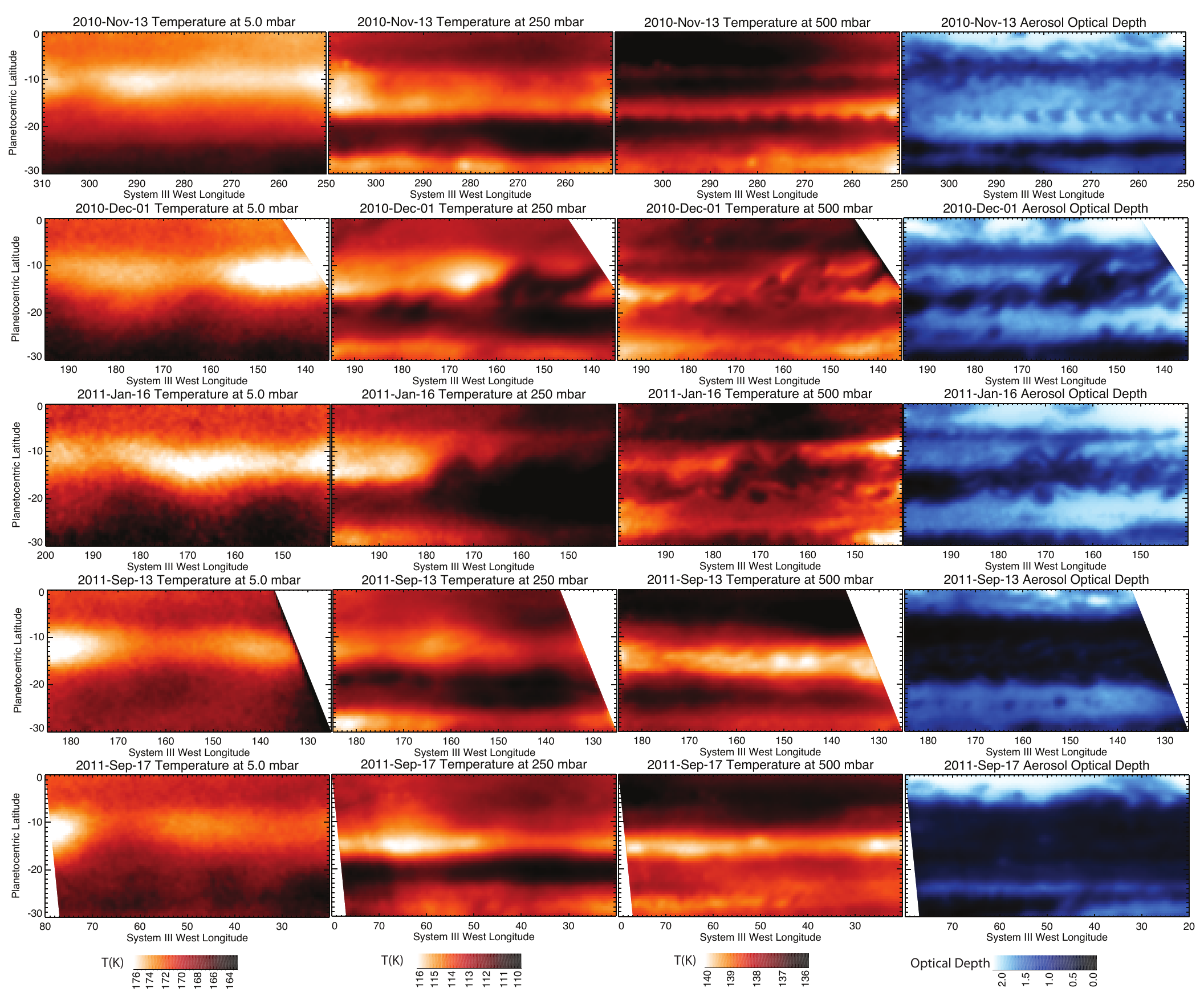}}
\caption{Retrieved temperature maps at three pressure levels (5 mbar, 250 mbar and 500 mbar) and aerosol opacity maps for five dates during the SEB revival - the faded state on November 13th 2010 (day 2), the early stages of the outbreak on December 1st 2010 (day 22), the evolving revival on January 16th 2010 (day 68), the revived state near the GRS on September 13th 2010 (day 308) and the revived state away from the GRS on September 18th 2011 (day 313).  Triangular white regions were not observed by all filters, so were omitted from the retrieval. }
\label{Tmap}
\end{centering}
\end{figure*}

Longitudinal variability is seen at 5 mbar (Fig. \ref{Tmap}, first column), with 2-6 K stratospheric temperature contrasts associated with the wave initiated by the revival (see Section \ref{stratwave}).  At 250 mbar, at the altitudes sensed by the Q-band filters, the faded SEB had a relatively uniform temperature of 113-114 K (Fig. \ref{Tmap}, second column).  Paper 1 concluded that changes to the thermal structure for $p<300$ mbar (i.e., in the stably-stratified regions to the upper troposphere) had been minimal during the fade.  However, Fig. \ref{Tmap} reveals that the convective outbreak created a dramatic temperature drop of 3-4 K over the active plumes at 250 mbar.  On December 1st 2010, this was coupled to a local maximum in temperature immediately to the northwest of the central branch, which can also be seen as elevated brightness in the individual filters in Fig. \ref{filters}.  While the central branch was expanding, the cooler 250-mbar temperatures (i.e., the tops of the convective cells) spread over a wide longitude range.  However, by the end of the revival (September 2011) these cool regions had been replaced by the typical 113-114 K temperatures, albeit with some longitudinal variability of $<1$ K amplitude.  

Changes to the temperature structure are more dramatic for $p>300$ mbar as we move into the more convectively-unstable regions of the troposphere. The November 2010 image reveals the structure that typified the faded state - narrow warm bands at the SEB(N) and SEB(S) separated by the cooler SEBZ (1.0-1.5 K temperature contrast at 500 mbar, see Fig. \ref{Tmap}, third column).  The active plumes disrupted this regular pattern in December 2010 and January 2011 and caused temperatures to drop, with localised warming occurring adjacent to the plumes.  In addition, there are more extended warm areas east and west of the plumes (corresponding to the `red' diffuse areas seen at 10.8 $\mu$m in Fig. \ref{visir_label}).  The 500-mbar temperature contrasts are small, being no more than 1-2 K between the coldest plume and the warmest lane.  It is the regions of subsidence surrounding the plumes that caused the warming and removal of the pale aerosols to reinvigorate the SEB's normal colours.  The cold plumes can no longer be seen in September 2011, when the warm SEB(S) band had reformed.  

\subsubsection{Aerosol variability}
The fourth column of Fig. \ref{Tmap} reveals that the revival resulted in a drop in the $\sim700$-mbar aerosol opacity by a factor of two \citep[assumed to be due to a layer of NH$_3$ ice particles, with optical depths determined at 10 $\mu$m using the refractive indices of][]{84martonchik}.  Fig. 8 of Paper 1 showed that the fade caused the cumulative optical depth to increase from 0.8 to 1.7, whereas Fig. \ref{Tmap} shows the reversal of that trend.  The November-13th aerosol map shows clear conditions over the SEB(N) near $7^\circ$S, mottled aerosol coverage over most of the SEB down to $16^\circ$S, an undulating wave pattern on the SEBs retrograde jet \citep{16rogers_wave} and a broad band of cloud opacity extending down to $\sim23^\circ$S (the South Tropical Zone).  Cloud opacity was dissipated quickly over the central branch of the revival (December 1st and January 16th aerosol maps).  Plumes with higher aerosol opacity than the surroundings can be seen at $13^\circ$S, $161^\circ$W (WS7) and $10^\circ$S, $151^\circ$W (WS3) on December 1st.  In addition, cloud-free patches can be seen on the SEB(S) westward of $165^\circ$W.  By the end of the revival, there is no contrast in the aerosol opacity within the SEB itself, just a broad cloud-free band between $6^\circ$S and $20^\circ$S with optical depths $<0.8$, consistent with pre-faded conditions.  It is this dissipation of aerosol opacity which permits the substantial increase in 5-$\mu$m radiance from the SEB observed in Fig. \ref{5um_maps}.

\subsubsection{Para-hydrogen}
Finally, we used IRTF/MIRSI imaging data at 18.75 and 24.5 $\mu$m to search for evidence that the upwelling associated with the revival modified the para-hydrogen fraction over the SEB.  A strong updraft would bring air with a low para-H$_2$ fraction upwards, whereas compensating subsidence would bring high-para-H$_2$ fraction air downwards, changing the S(0) absorption and the ratio of the brightness at these two wavelengths.  We compared SEB radiances pre-revival (November 4th 2010), mid-revival (December 5th 2010) and post-revival (August 30th 2011), and found no changes to the relative brightnesses of these two filters, beyond the temperature changes discussed above.  

\section{Discussion}
\label{discuss}

\subsection{Key properties of the SEB revival}

The 2009-2011 fade and revival cycle was the first to be observed at high spatial resolution in the thermal infrared, allowing us to understand the temperature changes (and implications for vertical motions) associated with the disturbance.  Additional insights into the 2010 revival were gained from the high-cadence tracking of the SEB revival by amateur observers, as summarised by \citet{11rogers_21, 11rogers_24, 15rogers, 16rogers}.  Furthermore, comparison to the historical record of SEB life cycles \citep{95rogers, 96sanchez} reveals regular patterns of activity and common features of the single-source region from which the central, southern and northern branches emanate.  The following list summarises the key observational characteristics identified in Section \ref{timeline}.  Taken together, this suggests that the SEB revival consisted of a series of convective plumes rising up through quiescent atmospheric layers and resulting in the removal of aerosols to reveal the regular coloration of the SEB.

\begin{enumerate}
\item \textbf{Origin: } The initial plume was closely associated with a cyclonic (formerly brown) barge that had been present in 2009-10 (Section \ref{seq0}).  However, no thermal or chemical changes were detected within this barge or the surrounding region that could have alerted us to the eruption of the first plume.  Prior to the eruption, the SEB appeared stagnant and quiescent, with white aerosols (a putative NH$_3$-ice haze) obscuring the usual colours of the belt.
\item \textbf{Upwelling: } The majority of plumes emanated from a single source and along a broad `leading edge' immediately east of the source region (Section \ref{seq1} and \ref{seq2}).  Plumes erupted every 2-6 days for a period of approximately 2 months, and had 4-22 day lifetimes during the early stages.  The several-day lifetimes of the plumes are consistent with $\sim3.5$-day lifetime of moist convective spots tracked by Cassini \citep{04li}. Plumes were cold-cored at their tops, white in colour and had a high aerosol opacity.  The plumes were bright in CH$_4$-band imaging at 0.89 $\mu$m, suggesting that they reached high altitudes into the lower stratosphere, particularly when they were new.  The rising convective plumes were localised in longitude and either (i) dredged white, pre-existing aerosols upwards from below the faded cloud deck; or (ii) brought moist NH$_3$-laden and H$_2$O-laden air upwards to condense in the rising plumes.  There was no evidence for latent heat injection from condensation in the $p<700$ mbar range, although this could have been hidden from 5-20 $\mu$m remote sensing by the overlying cold, divergent region at the top of the plumes.  
\item \textbf{Subsidence: } The cells containing the plumes were bounded by sharply-defined warm lanes, elongated from southwest to northeast, which were dark at visible wavelengths (Section \ref{seq1} and \ref{seq2}).  Additionally, the cells were flanked by diffuse warm regions to the west (e.g., Fig. \ref{visir_label} and \ref{filters}).  The warmth was likely due to large-scale subsidence surrounding the convective plumes, and caused clearing of 700-mbar aerosol opacity (sensed at 8.6 $\mu$m) and 2-3 bar opacity (5 $\mu$m) in the narrow lanes, potentially via sublimation of the putative NH$_3$-ice aerosols that had caused the SEB whitening in the first place.  
\item \textbf{Expansion of the revival: } The aerosol-free regions surrounding the plumes expanded in longitude as the leading edge of the central branch moved east from the source region.  Northern and southern branches formed as a chain of cloud-free patches moved west along the retrograde SEBs and north on the prograde SEBn jet (Section \ref{seq3}).  The full SEB was revived (i.e., dark brown in colour) $\sim200$ days after the onset of the disturbance.
\item \textbf{Re-established convection: } Normal convective activity northwest of the GRS resumed $316$ days after the disturbance outbreak (Section \ref{seq4}).  This resumption of activity occurred 838 days (2 years and 3.5 months) after the last reported activity in this region in June 2009 (Paper 1).  When convective rifting was absent, a cool SEBZ separated the warmer SEB(S) and SEB(N), whereas conditions were generally warmer to the west of the GRS when rifting was active.
\item \textbf{Troposphere-stratosphere coupling: } The tropospheric convection served as a source of vertically-propagating waves that interacted with the atmospheric motions in the mid-stratosphere, manifesting as a zonal thermal wave of $20-30^\circ$ longitudinal wavelength that was observed a few months after the start of the outbreak (Section \ref{stratwave}).
\end{enumerate}

\subsubsection{Comparison to Saturn's storm}
Coincidentally, a storm eruption of similar vigour was underway at the same time on Saturn, as observed by the Cassini spacecraft and ground-based observers \citep{11sanchez, 11fischer, 11fletcher_storm, 12fletcher, 12sanchez, 13sayanagi, 14achterberg}.  The similarities in the eruption pattern are striking, even though the Saturn storm may have been from a single localised source rather than multiple plumes.  The Saturnian plume erupted as a bright white spot in a relatively quiescent region of the northern hemisphere, near the peak of a retrograde jet and possibly associated with a cyclonic atmospheric feature \citep[the `string of pearls',][]{14sayanagi}.  The plume was cold-cored at its top with no detectable evidence of latent heating in the upper troposphere, and surrounded by regions warmed by large-scale subsidence \citep{11fletcher_storm}.  Sustained convection (and lightning strikes) were detectable over several months, and we speculate that the SEB revival could have caused similarly vigorous lighting activity on Jupiter.  The powerful convection radiated waves into the stratosphere, but rather than produce a wave pattern as on Jupiter, these waves produced a large, warm stratospheric anticyclone that persisted for almost three years \citep{12fletcher}. Finally, the aftermath of the storm left a warm, cloud-free band that persisted for several years after the outbreak \citep{14achterberg, 14fletcher_texes}, readily visible as a dearth of ammonia opacity at microwave wavelengths \citep{13janssen, 13laraia}.  

\subsubsection{Comparison to NTB plumes}
Further comparisons can be drawn with convective plumes erupting episodically on Jupiter's prograde jet on the southern edge of the North Temperate Belt \citep[the NTBs jet at $23.5^\circ$N, see Chapter 7 of][]{95rogers}.  The three most recent NTBs plume eruptions (1975, 1990 and 2007) \citep{91sanchez, 08sanchez}, featured two large plumes and left a turbulent and reddened NTB in their wake.  Modelling by \citet{08sanchez} suggested that a cluster of updraft cells, rising from the water cloud at 5-7 bar to the tropopause level, could be creating fresh, white aerosols in the plumes themselves.  However, they suggested that a red chromophore was being produced by an increased aerosol density in the NTB, counter to our explanation for the revival of the SEB brown coloration (subsidence and aerosol clearing).  

\subsection{Moist Convection on Jupiter}
Earth's troposphere, particularly in the rising branches of the Hadley cell, is characterised by narrow, rising air currents forming towering cumuliform clouds that are surrounded by broad, dry downdrafts \citep{00ingersoll}.  The primary characteristic of Jupiter's SEB is that it is warmer than the neighbouring zones in the upper troposphere, suggesting that the net effect of the vertical motion is adiabatic heating from the dry downdrafts \citep[e.g.,][]{86gierasch}.  The SEB revival plumes are likely to be moist convective storms (cumulonimbus clouds) rising against these prevailing downdrafts, using energy provided by latent heating from water condensation \citep[e.g.,][]{70barcilon,72kuiper,86stoker} to rise from the water cloud base \citep[4-6 bar, depending on Jupiter's bulk oxygen content,][]{14sugiyama} up to the tropopause, a vertical distance of $\sim100$ km.  \citet{86stoker} showed that, for a solar-composition atmosphere, NH$_3$ condensation could provide only a small fraction of the energy required to accelerate jovian equatorial plumes and that NH$_3$-driven plumes would quickly decelerate in the stably-stratified upper troposphere.  H$_2$O is therefore a better candidate energy source for vigorous cumulonimbus production on Jupiter, owing to its higher abundance and larger latent heat content than NH$_3$ \citep[see also][]{14sugiyama}.  Furthermore, the SEB plumes must rise from deeper than the NH$_3$ condensation altitude to explain the 5-$\mu$m observations (sensing the 2-3 bar level), although we stress that these data cannot definitively detect condensed H$_2$O in the plumes to confirm this hypothesis.  

\begin{figure*}
\begin{centering}
\centerline{\includegraphics[angle=0,scale=.50]{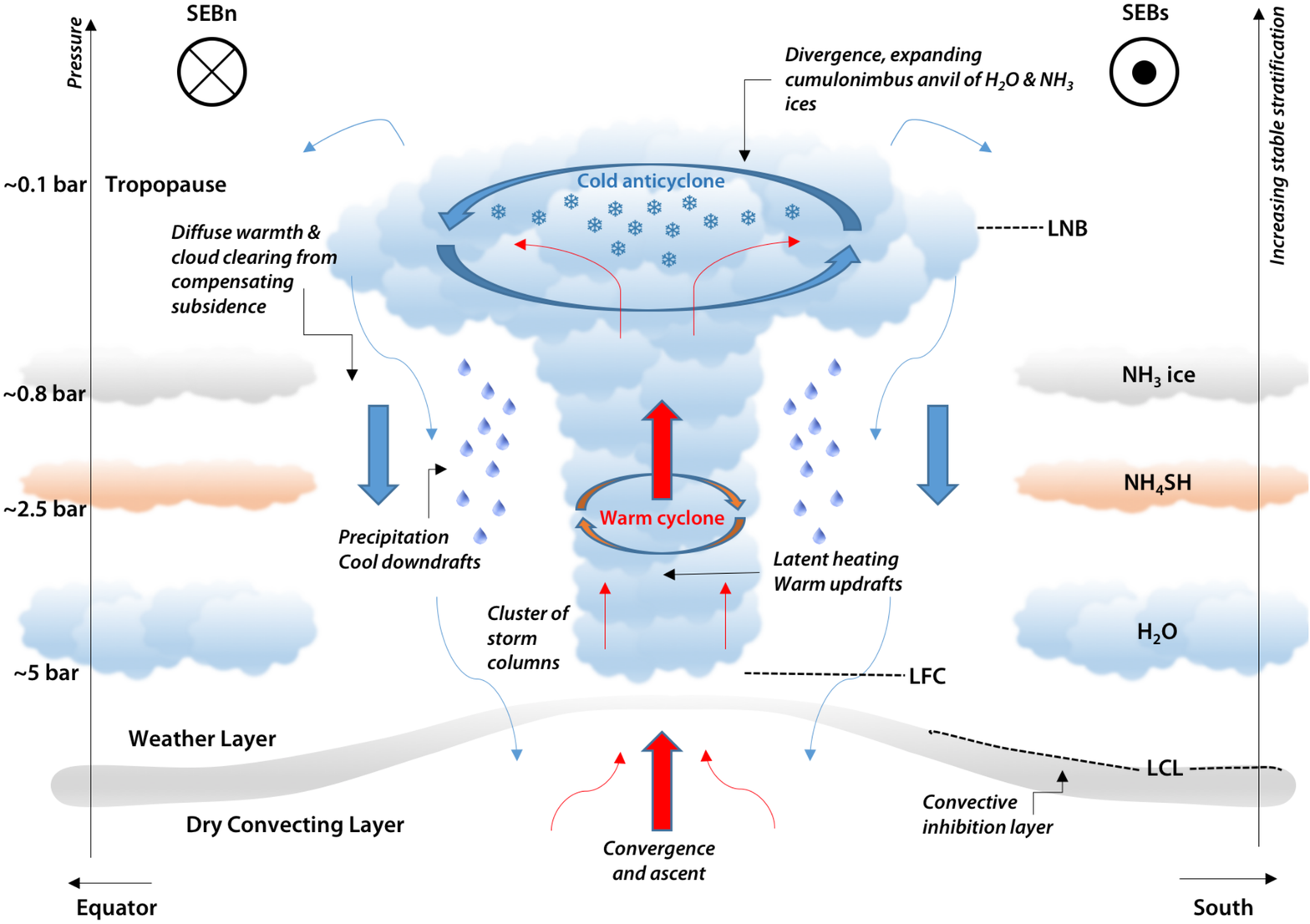}}
\caption{Diagram representing motions associated with a moist convective plume on Jupiter, following \citet{02hueso, 04houze, 16thomson}.  Latitudes increase from left to right, from the equator to the southern hemisphere.  The LCL (lifting condensation layer), LFC (level of free convection), LNB (level of neutral buoyancy) and CIN (convective inhibition layer) are all noted, as well as approximate altitudes for the three primary cloud decks.  Horizontal cyclonic and anticyclonic motions in mesoscale vortices, and vertical upwelling and subsidence creating plumes and lanes, are represented by arrows on the diagram.  All locations are approximate and the diagram is not to scale. }
\label{convection}
\end{centering}
\end{figure*}

Previous studies have suggested that moist convection is integral to maintaining Jupiter's banded structure \citep[e.g.,][]{70barcilon, 00ingersoll, 02hueso, 04ingersoll, 05showman, 14sugiyama, 16thomson} and the vertical transport of a substantial fraction of Jupiter's 5.7 W/m$^2$ of internal heat \citep{00gierasch}.  Given the large scales of convective plumes \citep[1000-5000 km,][]{00gierasch, 02hueso}, multi-day lifetimes, and the detection of lightning activity by Galileo, \citet{00gierasch} likened jovian plumes to \textit{mesoscale convective systems} (MCSs), or clusters of thunderstorms observed on Earth \citep[see detailed reviews in][]{93houze,04houze,06wallace,15thomson_phd}.  Indeed, single-cell storms have lifetimes on the order of hours \citep{01hueso}, whereas the multi-cell systems studied by \citet{02hueso} and \citet{14sugiyama} last for several days, more consistent with the observed lifetimes of convective spots on Jupiter \citep{04li} and the plumes reported here (Section \ref{seq1}).  In this section, we propose that many of the observed characteristics of the SEB revival can be understood via analogy to these terrestrial MCSs, albeit in the absence of a solid lower boundary.  The morphology of the SEB plumes resembles simulations of storm clusters advected by the background SEB windfield by \citet{02hueso}.  This analogy provides qualitative explanations for why the plumes cluster in one place, how they clear the atmosphere of condensates and volatiles, and the relationship between the storm eruption and cyclonicity.  A diagram representing the key features of an SEB revival plume is shown in Fig. \ref{convection}.  

\subsection{Vertical structure of convective complexes}

Fig. \ref{convection} subdivides Jupiter into a deep, dry-convecting region (unsaturated) and the moist-convecting `weather layer'.  A parcel of air rising from the deep atmosphere cools at the dry adiabatic lapse rate and becomes saturated at the \textit{lifting condensation level} (LCL).  Each cloud-forming molecule on Jupiter (H$_2$O, H$_2$S, NH$_4$SH, NH$_3$) has its own LCL, with the water cloud base representing the base of the weather layer in Fig. \ref{convection}.   The high molecular weight of moist jovian air, coupled with latent heat release from condensation that warms the local atmosphere to decrease the density above the LCL, should create a stably-stratified inversion between the cloud base and the overlying layers \citep[e.g.,][]{98showman, 14sugiyama, 15li}.  This stable layer would normally inhibit convective motions \citep[in practice, the relative humidity of the environment needs to be high enough to permit buoyancy,][]{86stoker}.  But if a parcel were forced to continue rising it would cool at the saturated adiabatic lapse rate until it reaches the \textit{level of free convection} (LFC).  Further motion above the LFC causes condensation and latent heating, and the upward motion continues unaided, driven by the buoyancy difference between the moist adiabat and the cooler, dry adiabat.  Latent heating continues until the plume reaches the \textit{level of neutral buoyancy} (LNB, near the stably-stratified tropopause), where the column begins to diverge and cool.  Volatiles that had been transported upwards in the column are fully condensed and precipitate out in downdrafts as liquid or ice, and the topmost cloud expands to create the broad anvils that are common to thunderstorms on Earth. 

An MCS on Earth consists of multiple moist convective storms of this nature, generally with newly-forming plumes on one edge, mature plumes in the centre and dissipating plumes on the other edge, with the direction determined by the background wind shear \citep{93houze, 04houze}.  Two common types of MCS are \textit{squall lines}, where a chain of thunderstorms form along a cold front, and \textit{mesoscale convective complexes} (MCCs), characterised as long-lasting ($>6$ hours to multiple days), quasi-circular storm systems with intense precipitation and very cold cloud tops seen in satellite imagery over areas exceeding $10^5$ km$^2$ \citep{93houze}.  The SEB plumes discussed here have a comparable lifetime (multiple days, Section \ref{seq1}), but larger area ($>10^6$ km$^2$, Section \ref{seq2}) than their terrestrial MCC counterparts.  It is the cold cloud top that we see in infrared images of the SEB revival plumes (Fig. \ref{convection}), but it is possible that the leading edge of the disturbance (where multiple additional plumes were observed) is analogous to a squall line.  MCCs develop in locations of weak background shear \citep{04houze}, similar to those found in the centre of the SEB ($dT/dy$, and therefore $du/dz$ via thermal wind balance, approaches zero in the centre of the belt). 

The large extent of MCCs means that the Coriolis force plays a role to generate a \textit{mesoscale convective vortex} (MCV):  the injection of energy to warm the low- and mid-levels results in cyclonic motion, whereas the divergence and cooling at the top level (i.e., the anvil cloud) results in anticyclonic motion \citep{94emanuel, 06wallace}.  On Earth, this MCV can persist for some time after the storm has abated.  The same geostrophic balance applies on Jupiter, such that the SEB plumes represent organised cumulus convection held together by a common circulation pattern.  The anticyclonic motion at the top of the white plume opposes the general cyclonic winds of the SEB, so once the source is exhausted the putative anticyclone loses coherence \citep{02hueso} and breaks into the complex eddies that characterise the brown revival region.

The plumes should display warm, cyclonic circulation at the mid-levels (Fig. \ref{convection}).  However, the 5-20 $\mu$m remote sensing presented here detected no evidence for such heating. Instead we see only the cold, anticyclonic upper levels of the cloud.  \citet{01hueso} investigated the structure of a single-cell convective cloud using a three-dimensional model, triggering convection via a 0.15-K perturbation at the base of the H$_2$O cloud.  \citet{02hueso} extended this to a cluster of convective columns being advected by the 2D windfield of the SEB.  Adopting their thermal structure for the resulting plume (a $\sim3$-K warming for $p>500$ mbar and a $\sim8$-K cooling near $p=200$ mbar), we explored how cold the upper, anticyclonic part of the plume would need to be to mask the warming at depth.  Unsurprisingly, the Q-band filters are the least sensitive to the deep warm core, and require only a moderately cold plume top ($\Delta T<1$ K) to hide the warm core.  N-band filters require cold tops with $\Delta T\sim2-4$ K to ensure that the warm core remains hidden, with the exception of the 8.6-$\mu$m filter.  Here even a cloud-top 10 K cooler than the surroundings cannot mask the warm core, and additional opacity due to aerosols is required.  Adding aerosol opacity reduces the size of the temperature anomaly required to reproduce the data, but unfortunately photometric imaging cannot be used to distinguish between temperature changes and aerosol opacity changes within the rising plumes.  In summary, the cold, cloudy plume tops of the MCC would successfully hide any warm-core features in the deeper troposphere from 5-20 $\mu$m remote sensing.

\textit{What about the subsiding branch of this circulation?}  On Earth, MCCs also have downdraft cells: upper-level downdrafts adjacent to the updraft cores, and lower-level downdrafts associated with intense precipitation \citep{93houze}.  In the latter case, droplets of rain or snow descend and evaporate, taking latent heat from the environment to cool the air and enhancing the negative buoyancy in the downdraft, eventually spreading out as a \textit{gust front} near the Earth's surface \citep{06wallace}.  The strength of the downdraft grows until it eventually cancels the upwelling, causing dissipation of the storm.  The SEB revival downdrafts consist of dry air from the upper troposphere, which warms along the dry adiabat as it descends due to compression.  The detectable warming above the cloud tops (i.e., in the 100-600 mbar region) creates both the diffuse emission surrounding the cells (Fig. \ref{visir_label}), and the more localised warming within the sharply-defined lanes.  This warming ultimately spread over the whole SEB, obliterating the cold upper-level temperatures associated with both the plumes and with the original cold SEBZ that characterised the fade (Section \ref{summary}). The warming contributes to the re-evaporation of the condensates (the obscuring white aerosols), and the descent of the dry air in cells is expected to cause further depletion of other tropospheric species - ammonia, phosphine and H$_2$S - which decrease in abundance with height, in addition to those that have already been precipitated.  This compensating subsidence may explain the observed NH$_3$ depletion to several bars over much of Jupiter \citep{01depater, 05showman}, as well as the observed dearth of NH$_3$ opacity in the aftermath of Saturn's springtime storm \citep{13janssen, 13laraia}.  After the downdrafts, the remaining atmosphere of the SEB would be unsaturated and stable.  

\subsection{Triggering moist convection}
Triggered convection occurs because of an accumulation of \textit{convective available potential energy} (CAPE) immediately below a cloud deck LCL.  The molecular weight gradient (saturated air being denser than dry air) creates a compositionally stable layer that inhibits convection \citep{00nakajima, 15li}.  Furthermore, the newly-formed upper cloud that was responsible for the fade (Paper 1) may absorb sunlight, further reducing the lapse rate and limiting convection in the faded SEB.  For the explosive convection to occur on Jupiter, air parcels must first be raised to the water LFC through the base of the weather layer \citep{86stoker}, which requires either forcing through the stable layer \citep[e.g., Fig. \ref{convection} and][]{14sugiyama} or erosion of the stable layer to allow convection \citep[e.g., ][]{15li}.  

On Earth, new cumulonimbus clouds may be formed by the gust front forcing warm air to rise above it to form new cells (explaining why clustering occurs near to the first convective plume), or due to instabilities occurring during the summer with warm air on the Earth's surface and cooler air aloft.  It is unclear how similar mechanisms would work on Jupiter to trigger multiple storm plumes near to a single source \citep[e.g.,][]{02hueso}.  However, \citet{14sugiyama} presented a mechanism to force parcels upwards during isolated convective events.  Their study suggested that, during `normal' jovian activity, quiescent periods between plumes corresponded to weak vertical motions above the stable regions associated with the NH$_3$ and NH$_4$SH LCLs.   As motions at these cloud decks intensify, downdrafts carry H$_2$O condensates down towards the water cloud base, where re-evaporation causes latent cooling.   These cooled airmasses could then penetrate through the water LFC and LCL and into the dry (i.e., unsaturated) convecting layers (Fig. \ref{convection}).  This subsidence must be balanced by parcels rising up from the unsaturated region, through the stable layer, delivering more water to the cloud base and becoming saturated to trigger the convection, hence generating the self-sustaining cumulonimbus storm system.  The sinking, evaporatively-cooled downdrafts tap into, and ultimately exhaust, the accumulated CAPE beneath the water cloud base.  


However, the model of \citet{14sugiyama} was pertinent to the `normal' state of the SEB, not the unusual conditions of the fade and revival.  During the revival, plumes were erupting every 2-6 days, but the period of quiescence before the revival was much longer than anything observed under `normal' conditions. This hints at a massive accumulation of CAPE during the faded state of the SEB, which required eruptions over $\sim100$ days to fully exhaust. But what changed to permit such accumulation, leading to the fade?  As in Paper 1, we speculate that some alteration of the deep flows around the GRS began to inhibit the usual quasi-continuous moist convection.  This was replaced by a slow, steady injection of fresh NH$_3$-rich air into the upper troposphere that thickened and whitened the pre-existing clouds \citep{12perezhoyos}.  But the sensitivity of the `normal' convection to the motions of the GRS remains to be explored via numerical simulations.  

The timescales for the SEB life cycle are intimately tied to the availability of water in the SEB, which acts as both the convective inhibitor during the fade and the source of latent heating during the revival \citep[e.g.,][]{01hueso, 02hueso, 14sugiyama, 15li}. However, \citet{14sugiyama} estimate that Jupiter's water abundances would need to be $20-30\times$ solar if their evaporative cooling model is the explanation for the long quiescence of the SEB.   If evaporation-powered downdrafts are not to blame, then the passage of some deep circulation pattern could have provided the trigger to release the accumulated CAPE (Section \ref{cyclone}).  An alternative explanation for the triggering comes from the study of Saturn's planetary-scale storm eruptions:  rather than invoking re-evaporation of condensates falling through the LCL, the mechanism of \citet{15li} erodes the stably-stratified layer via radiative cooling from the top of the atmosphere to the LCL.  The cooling proceeds downward, eventually increasing the density just above the water LCL until the molecular weight gradient is removed, permitting the release of the stored CAPE.  Once the CAPE is exhausted, the condensates precipitate out to re-establish the stable gradient.  Provided the water abundance is above some critical threshold to permit the formation of the inhibiting layer, the timescale for the radiative heat loss then drives the periodicity of the outbursts ($\sim70$ years in the Saturnian case).  Intriguingly, \citet{15li} suggest that the present estimate of $\sim5\times$ solar water enrichments on Jupiter is insufficient for this triggering mechanism to work, and that more continuous convection (such as that seen during the `normal' state of the SEB) should prevail.  Nevertheless, convective inhibition is clearly at work during Jupiter's faded state, and the timescales of the event are yet to be explained - 2-14 years between fades; 0.5-3 years of convective inhibition near the GRS ($\sim800$ days in 2009-10), and $\sim100$ days of outbursts during the SEB revival.

\subsection{Convection in cyclonic regions}
\label{cyclone}
Jupiter's belts are regions of cyclonic shear with frequent folded, filamentary and turbulent cloud structures \citep{95rogers, 04ingersoll}.  The belts are the sites of rare detections of spectroscopically-identifiable ammonia ice \citep{02baines} and vigorous lightning storm activity associated with moist convection \citep{92borucki, 99little, 00gierasch, 03porco}.  The mesoscale convective vortices produced in the low- to mid-tropospheric levels by the MCCs of Jupiter's SEB revival are also expected to be cyclonic.  On Earth, the MCVs can persist for days after the storm, and can serve to regenerate and amplify new convection.  Furthermore, we find that the initial SEB disturbance and subsequent central branch plumes erupted from the location of a cyclonic barge, and also in 2007 during the weaker SEB revival.  Saturn's 2010 eruption may also have occurred near to the cyclonic vortices of the `string of pearls.' \citep{14sayanagi}.  Finally, moist convection has been previously observed within jovian cyclones by Voyager 2 \citep{79smith, 15thomson_phd}\footnote{See \mbox{http://photojournal.jpl.nasa.gov/catalog/PIA02257}}.  So what is the connection between cyclonicity in the belts and barges and the promotion of moist convective activity?

\citet{89dowling_dps} noted that cyclones were regions of low pressure, causing isentropes (lines of constant potential temperature) to be depressed in upper layers and to rise up in the deeper layers.  They theorised that cyclones cause upward displacement near the water clouds that should be sufficient to trigger moist convection in regions of modest relative humidity.  More recently, \citet{16thomson} suggested that, provided the weather layer jets decay with depth into the dry convecting layer, thermal wind balance would cause isentropes to tilt upwards beneath belts, meaning that a closer connection between the water LCL and the LFC would exist beneath the belts \citep[see Fig. 1 of][and Fig. \ref{convection}]{16thomson}.  The stably-stratified region at the base of the weather layer would therefore be more extensive beneath zones than belts, qualitatively supported by the observation that the most vigorous SEB revival eruptions occurred near the centre of the belt where the interface layer would be shallowest, rather than at the edges.  Cyclonic regions could therefore provide access to the abyssal layers beneath the weather layer, promoting explosive moist convective outbursts such as those seen during the SEB revival.

\section{Conclusions}
\label{conclusion}

The 2010-2011 revival of Jupiter's South Equatorial Belt (SEB) was scrutinised at thermal-infrared (5-20 $\mu$m) wavelengths for the first time, allowing us to determine the temperature changes associated with the dissipation of the white aerosols responsible for the 2009-2010 fade \citep{11fletcher_fade, 12perezhoyos}.  A coordinated campaign of photometric imaging was necessary to track this eruptive storm system over twelve months, utilising the VLT and Gemini-South in Chile; the IRTF, Subaru and Gemini-North in Hawaii; in tandem with amateur observers from around the globe.  A comprehensive timeline for the revival is shown in Table \ref{tab:timeline} and described in Section \ref{timeline}, using spectral inversion to map variations in the temperatures and aerosol opacity in the upper troposphere and stratosphere.  The key findings are discussed in Section \ref{discuss}, which demonstrates that an analogy to moist convection in terrestrial mesoscale convective systems \citep[squall lines and thunderstorm clusters,][]{00gierasch} can be used to explain some of the features of the revival.  Thermal-IR imaging senses the cold, divergent and putatively anticyclonic upper cloud layers of the plumes (akin to the anvil tops of cumulonimbus clouds), which may reside above (undetected) cyclonic vortices at mid- and lower-levels of the weather layer, warmed by latent heating from water condensation.   Subsidence surrounding the plumes is initially localised in cells with dark, cloud-free lanes on their western boundaries.  This sinking leads to compressional heating, particularly to the northwest and southeast of the central branch, re-evaporating the putative NH$_3$ ices that were responsible for the fade.  The unsaturated, dry air is eventually redistributed around the planet in the central, northern and southern branches until the whole SEB has regained its typical brown appearance.  The tropospheric upwelling from the central branch was sufficiently vigorous to excite longitudinal thermal waves at stratospheric altitudes, proving a direct connection between tropospheric convection and stratospheric wave phenomena. 

Moist convection, powered by a jovian water cycle (latent heating to provide buoyancy for rising plumes; evaporative cooling to provide downdrafts into the deep, dry-convecting layer) appears to be central to the life cycle of the SEB.  When quasi-continuous convection is active in the turbulent, filamentary region northwest of the Great Red Spot, the convective available potential energy (CAPE) is only stored for a limited time, on the order of days.  But when this convection shuts down, for as-yet unknown reasons, the CAPE can accumulate until an explosive release is triggered.  The absence of convection during the fade permits the formation of a cool SEBZ (zone-like band in the SEB centre) and the thickening of the existing clouds (e.g., adding fresh NH$_3$ ice) to cause the fade \citep{11fletcher_fade, 12perezhoyos}.  The presence of this new aerosol layer forms a stable layer in the upper troposphere, reinforcing the quiescent appearance of the whitened belt.  The timescale between the cessation of moist convection and the triggered release is not understood, but the trigger appears to be associated with residual cyclonic motions (and warmer temperatures detected at 9-13 $\mu$m) associated with a formerly brown barge that had been a characteristic feature of the faded SEB.  The bias of moist convection to cyclonic belts (e.g., thunderstorm activity) and cyclones has been previously observed on Jupiter \citep{00ingersoll}, and may be related to a shallower, thinner interface between the weather layer and the deeper, dry-convecting interior below cyclonic regions \citep[e.g.,][]{89dowling, 16thomson}.  

Once triggered, the SEB revival became self-sustaining and spread from the original outbreak region.  Downdrafts associated with the narrow rising plumes could penetrate back to the water cloud, triggering more lifting of air parcels to the level of free convection and explaining why the majority of plumes erupted from one source (analogous to a mesoscale convective complex) and along a broad line to the east (the leading edge, analogous to a squall line).  The background shear of the SEB, bordered by the retrograde SEBs jet to the south and the prograde SEBn jet to the north, controlled the redistribution of warmed, cloud-free and unsaturated air to revive the SEB(S) and SEB(N) in a westward-moving southern branch and an eastward-moving northern branch, following patterns recorded in previous SEB revivals \citep{95rogers, 96sanchez} and numerical simulations \citep{02hueso}.  The cool SEBZ and any evidence of the cold, anticyclonic plume tops were eventually replaced by the warm, aerosol-free state, and the `normal' quasi-continuous convection northwest of the GRS began again $\sim316$ days after the outbreak, and $\sim838$ days since its last appearance.  

Although the thermal-IR observations of the revival have highlighted the importance of moist convection in large-scale, organised complexes, progress in understanding the timescales associated with the SEB life cycle (2-4 years between cycles, 100 days of convective eruptions, 0.5-3 years without `normal' convective activity near the GRS) will rely on future numerical simulation.  Models of triggered moist convection with evaporation-driven downdrafts through the cloud base \citep{14sugiyama} or radiative cooling to erode the stably-stratified layers causing convective inhibition \citep{15li} show considerable promise.  On the observational side, photometric imaging can provide crude estimates of temperature changes, but it is insufficient to reveal the details of the redistribution of chemicals occurring during the SEB revival.  Rising plumes might be expected to loft phosphine, low-para-H$_2$ air, spectroscopically-identifiable NH$_3$ and H$_2$O ices into the upper troposphere.  Similarly, subsiding downdrafts should deplete NH$_3$, H$_2$O and other volatiles in the wake of the storm.  Spectroscopic mapping is the next natural step to understand the distribution of these gases during a revival, and will be proposed whenever the next opportunity arises.  We eagerly await the results of the Juno spacecraft's microwave radiometry in 2016-17, which may permit new understandings of the interaction between the zonal flows and the GRS at levels deeper than those accessible to thermal-IR remote sensing.  Finally, future observing campaigns for SEB revivals should focus on cyclonic regions like the barges as potential sources of the outbreaks.

\section*{Acknowledgments}

Fletcher was supported by a Royal Society Research Fellowship at the University of Leicester, Giles was supported by a Royal Society research grant at the University of Oxford.  The UK authors acknowledge the support of the Science and Technology Facilities Council (STFC).  A portion of this work was performed by Orton and Payne at the Jet Propulsion Laboratory, California Institute of Technology, under a contract with NASA.  This research used the ALICE High Performance Computing Facility at the University of Leicester.  We are extremely grateful for the combined efforts of the numerous amateur observers (including those listed in the figure captions) for sharing their data, and for the JUPOS software developed by Grischa Hahn and Hans-J{\"o}rg Mettig to reproject the visible-light data.  

This investigation was partially based on thermal-infrared observations acquired at (i) the ESO Very Large Telescope Paranal UT3/Melipal Observatory using Directors Discretionary Time (program ID 286.C-5009) and regular service time (program ID 087.C-0024); (ii) the Subaru Telescope, which is operated by the National Astronomical Observatory of Japan (program ID O11154); (iii) NASA's Infrared Telescope Facility,  which is operated by the University of Hawaii under contract NNH14CK55B with the National Aeronautics and Space Administration (program IDs  2010B010, 2011A010, 2011B027); and (iv) observations obtained at the Gemini Observatory (program IDs GN-2010B-DD-3, GS-2010B-Q-8 and GS-2011A-Q-11), which is operated by the Association of Universities for Research in Astronomy, Inc., under a cooperative agreement with the NSF on behalf of the Gemini partnership: the National Science Foundation (United States), the National Research Council (Canada), CONICYT (Chile), Ministerio de Ciencia, Tecnolog\'{i}a e Innovaci\'{o}n Productiva (Argentina), and Minist\'{e}rio da Ci\^{e}ncia, Tecnologia e Inova\c{c}\~{a}o (Brazil).  We wish to recognise and acknowledge the very significant cultural role and reverence that the summit of Mauna Kea has always had within the indigenous Hawaiian community.  We are most fortunate to have the opportunity to conduct observations from this mountain.

\section*{References}
\bibliographystyle{elsarticle-harv}
\bibliography{references}







\end{document}